\documentstyle[aps,prc,preprint]{revtex}   

\begin{document}

\preprint{
DPNU-98-15,
TMUP-HEL-9804,
DUKE-TH-98-159,
DOE/ER/40561-7-INT98
}
\draft
\title{  Anomaly Induced Domain Formation of \\
         Disoriented Chiral Condensates }
\author{Masayuki Asakawa\thanks{yuki@nuc-th.phys.nagoya-u.ac.jp}\\
        Department of Physics, School of Science, Nagoya University\\
        Nagoya, 464-8602, Japan \\
        and\\
        Institute for Nuclear Theory, University of Washington\\
        Seattle, WA 98195-1550, U.S.A\\
        Hisakazu Minakata\thanks{minakata@phys.metro-u.ac.jp}\\
        Department of Physics, Tokyo Metropolitan University  \\
        Minami-Osawa, Hachioji, Tokyo 192-0397, Japan \\
        Berndt M\"uller\thanks{muller@phy.duke.edu}\\
        Department of Physics, Duke University\\
        Durham, NC 27708-0305, U.S.A.\\
        }
\date{May 1, 1998}

\maketitle
\begin{abstract}
We discuss the effect of chiral anomaly as a possible mechanism for 
triggering formation of domains of disoriented chiral condensate (DCC)
in relativistic heavy ion collisions. The anomalous
$\pi^0 \rightarrow 2 \gamma$ 
coupling and the strong, Lorentz contracted electromagnetic fields of 
the heavy ions combine to produce the ``anomaly kick'' to the field 
configuration of the neutral pion field. We implement the effect of 
anomaly kick in our numerical simulation of the linear sigma model 
in a schematic way which preserves its characteristic features:
the effect is coherent over a large region of space but is opposite 
in sign above and below the ion scattering plane. 
We demonstrate by detailed simulations with longitudinal expansion 
that the DCC domain formation is dramatically enhanced by the anomaly 
kick in spite of its small absolute magnitude. We examine the behavior of
various physical quantities such as pion fields, the axial vector 
currents, and their correlation functions. Our results also provide
useful insight into the mechanism and properties of DCC domain formation,
in general. Finally, we discuss some experimental observables which can 
signal the anomaly induced formation of DCC.
\end{abstract}



\section{Introduction}
\label{sec_introd}
One expects, on account of universality arguments, that quantum
chromodynamics (QCD) with two massless quark flavors exhibits a 
second-order phase transition between a low-temperature phase, 
which shows spontaneous chiral symmetry breaking, and a chirally
symmetric high-temperature phase \cite{pw84,rw92}.
The fate of this transition for the physical values of the
$u$- and $d$-quark masses and a third semi-light quark flavor ($s$),
is currently under intense investigation by means of numerical
simulations of the lattice gauge theory. If the transition exists
in the real world, e.g., in intermediate states in heavy ion
collisions, an interesting phenomenon can occur:
Large coherent domains of pion fields form due to the long range
correlations associated with the second order phase transition.
Unfortunately, it is unlikely that the second order transition
persists with finite quark masses, as it is easily destroyed by
a weak external magnetic field in the case of magnetization,
and we do not expect that large domains form in an equilibrium
situation \cite{rw92}.

However, it was argued by Rajagopal and Wilczek \cite{rw92}
that domains may be formed in energetic collisions where hot 
regions experience subsequent non-equilibrium evolution.  
They proposed an idealized quench approximation to 
model this non-equilibrium scenario \cite{rw93}. It provides
a concrete realization for the formation mechanism of large
chirally misaligned domains, the disoriented chiral condensates
(DCCs). Such chirally misaligned coherent pion field domains
have been discussed by many authors \cite{raja97}
after the pioneering works \cite{anselm,bjorken1,blaizot1}
in the context of large neutral-charged pion fluctuations 
(Centauro events) that may have been seen in cosmic ray experiments
\cite{centauro}.

The central question in the field, i.e., whether the DCC forms in high
energy hadronic collisions, has been discussed extensively.
We summarize these discussions below, and then we point out that
the chiral anomaly combined with the environment of
relativistic heavy ion collisions greatly enhances the possibility of
formation of DCC domains.

The possibility of the formation of large DCC domains is investigated 
by means of numerical simulation \cite{rw93,ggp94,ahw95} 
using the linear sigma model \cite{gl} as a low energy 
effective theory of QCD. Among the most elaborate of these simulations,
one with longitudinal expansion \cite{ahw95} indicates that large 
spatial domains of DCC can, indeed, develop for appropriate 
initial conditions.

Disoriented chiral domains can only form when a region of space is
cooled down rapidly and chiral symmetry gets spontaneously broken
sufficiently fast, so that the chiral order parameter
$\langle \bar{q} q \rangle$ cannot adiabatically follow the
shifting minimum of the effective potential.
This is the reason for the quench scenario.
To realize it, the hot debris 
formed in high energy hadronic collisions must somehow be cooled
down quite rapidly. A one-dimensional expansion does not appear
to be sufficiently fast \cite{ahw95,rand96}, as is expected from 
the early calculation using the hydrodynamical model by Bjorken 
\cite{bjorken2}. Therefore, a rapid three-dimensional expansion 
would be required \cite{rand96,gm94} to drive the chiral field 
far enough out of equilibrium.

If the relics of such domains were observed in experiments where
matter is heated above $T_c$, e.g., in relativistic heavy ion
collisions, they would provide evidence for the existence for the
chiral phase transition in QCD in non-equilibrium environments.
Because the domains decay into coherent multi-pion states,
they are predicted to reveal themselves in highly characteristic
pionic observables. The unusual distribution of the pion charge
ratio $R = N_{\pi^0}/N_{\pi}$ has been discussed extensively in
the literature \cite{rw93,anselm,bjorken1,blaizot1,ggm93},
and it is now the subject of active experimental investigations
\cite{minimax,WA98}.
However, it has been shown that the charge ratio is useful as
a sensitive signal for DCC formation only if a very small number
of independent domains are formed \cite{al96,rt97}, indicating
the need for more advanced analyses. The possibilities include
the wavelet analysis \cite{rt97} and multiparticle correlations
\cite{ggm93,minimax}. If the DCC can be described by a
squeezed state \cite{kogan} of the pion fields, as expected by
the parametric resonance mechanism \cite{mm95}, some peculiar
two pion correlations are expected \cite{hm97}.

We demonstrate in this paper that the Adler-Bell-Jackiw chiral U(1)
anomaly \cite{ABJ} expressed as the Wess-Zumino term \cite{WZW}
in the linear sigma model does affect the chiral orientation of
pion fields, and thereby enhances the DCC domain formation. 
As will be explained in the following sections, we formulate 
the effect of the anomaly as a ``kick'' to the neutral pion fields 
imparted by the electromagnetic fields of colliding relativistic 
heavy ions \cite{mm96}, and then implement it into a 
numerical code simulating the linear sigma model \cite{ahw95}.
We show by examining the various physical quantities such as pion
fields, the axial vector currents, and their correlation functions that
the anomaly kick, even with a tiny amplitude, acts as an effective
trigger for the formation of the DCC domains.

Our manuscript is organized as follows. In Section II we will review
relevant aspects of the dynamics of the axial anomaly interaction in
relativistic nuclear collisions. In Section III we discuss various
issues of our numerical calculations, in particular, the initial
conditions for the chiral fields and the method used to solve the
field equations. In Section IV we present and discuss our numerical
results: the effect of the anomaly kick on domain formation, the
space-time evolution of the vector and axial currents, the evolution
of the pion correlation functions, and the pion density distributions.
We summarize our results and make some suggestions for future
experiments in Section V.

\section{Effect of Chiral Anomaly in Domain Formation in DCC; 
Anomaly Kick}
\label{sec_kick}
We review the discussion in Ref. \cite{mm96}, in which two of us 
raised the question of possible effects of the chiral anomaly in 
triggering DCC formation in heavy ion collisions. There
the influence of the strong electromagnetic fields of colliding 
ions on the isospin orientations of fields was discussed in the 
framework of the linear sigma model. This is a natural question 
to ask because electromagnetism breaks isospin. However,
an explicit one-loop computation of the effective potential in the 
linear sigma model reveals that the isospin orientation of the ground state 
is not affected by the uniform background electromagnetic fields. 
Moreover, it can be argued on the basis of symmetry considerations that 
this result remains valid for any spatially and temporally varying 
background electromagnetic fields. Therefore, it was apparent that 
no effect is expected at the one-loop level. 

However,  the above argument contains an interesting loophole.
It is the presence of the Wess-Zumino term which represents the 
effect of the chiral anomaly in the effective theory of QCD. It induces 
the $\pi^0 \rightarrow 2 \gamma$ coupling in the Lagrangian of the linear 
sigma model as
\begin{equation}
{\cal L} = 
{1\over 2} \partial_{\mu}\vec\phi \partial^{\mu}\vec\phi -
{\lambda\over 4} \left(\vert\vec\phi\vert^2 - f_{\pi}^2\right)^2 +
H \sigma + {\alpha\over\pi f_\pi} \vec{E}\!\cdot\!\vec{H} \pi_3 
\label{lagrangian},
\end{equation}
where $\vec\phi = (\sigma,\vec\pi)$. 
The ground state of the chiral field therefore becomes sensitive to 
the background electromagnetic fields through the anomaly term.  

Then, what is the effect of the anomaly term on the chiral orientation 
of the pion fields, and in particular on the formation of DCC domains?
To answer the question we first note that the equation 
of motion of the $\pi_3$ field is altered to
\begin{equation}
\dot\Pi_3 = \nabla^2 \pi_3
-\lambda \left(\vert\vec\phi\vert^2 - f_{\pi}^2\right) \pi_3
+ {\alpha\over\pi f_\pi} \vec{E}\!\cdot\!\vec{H},
\end{equation}
where $\Pi_3$ denotes the conjugate field of $\pi_3$.
Then, $\Pi_3$ is affected by the strong electromagnetic fields of 
heavy ions during a collision by the amount 
\begin{equation}
{\mit\Delta}\Pi_3 =
{\alpha\over\pi f_\pi} \!\int\! \vec{E}\!\cdot\!\vec{H} dt, 
\end{equation}
where we have used the fact that the displacement of $\vec{\phi}$
is negligible during the collision as its duration is
quite short at RHIC and LHC \cite{mm96}.
If we use the point-nucleus approximation, $\vec{E}\!\cdot\!\vec{H}$ 
can be easily evaluated by computing the Lienard-Wiechert potential. 
One obtains
\begin{eqnarray}
\vec E\!\cdot\!\vec B &=& {2Z^2e^2\over (4\pi)^2 M}\; {\gamma^2 \over
R_1^3R_2^3} (\vec r\!\cdot\! \vec L) \label{EdotB}, \\
R_{1,2} &=& \sqrt{\gamma^2 (z\mp vt)^2+ \left( \vec r_{\perp}
\mp {\vec b\over 2}\right)^2},
\end{eqnarray}
where we have assumed a collision of identical nuclei
with mass $M$ and charge $Z$, and 
$\gamma$ denotes the Lorentz factor in the center of mass frame, and 
$\vec L=\vec b\!\times\! M\vec v$ with
impact parameter $\vec b$. 

The physical picture resulting from the above computation is 
as follows. The colliding two heavy ions have Lorentz contracted 
electromagnetic fields which produce a non-vanishing 
$\vec{E}\!\cdot\!\vec{H}$ for scatterings with angular momenta higher 
than $s$-wave. During the collisions the interaction region of 
space-time is affected by the anomaly term prior to the DCC formation. 
Since the collision time scale $R/\gamma$ is much shorter 
than that of the DCC domains that may be formed
in relativistic collisions, it is conceivable that it primarily 
affects the initial condition of the chiral fields. 
To visualize this effect of the anomaly term, we have introduced  
a term ``anomaly kick'' and described the effects
imparted by the electromagnetic fields on the initial field
configuration.

The anomaly kick has a distinct characteristic feature which 
should be important in detecting its experimental signature. 
The $\Pi_3$ field is kicked 
toward opposite directions in isospin space in the upper and lower 
half-spaces off the scattering plane, as seen in (\ref{EdotB}). 
A rough estimate in Ref. \cite{mm96}, however, indicates that 
the magnitude of the kick is small, 
${\mit\Delta}\Pi_3 \sim 0.1 m_{\pi}^{-2}$. 
On the other hand, it is coherent over nuclear dimensions in 
each half-space. Therefore, while it was argued in Ref. \cite{mm96} 
that the anomaly kick might produce interesting effects, it was 
difficult to draw definite conclusions on whether the anomaly effect 
enhances the domain formation of DCC.

In this paper we implement the kick to the conjugate field 
of the neutral pion field as an initial condition of the numerical 
simulation code of DCC domain formation using the linear sigma 
model \cite{ahw95}. 

We now elaborate the above estimate of ${\mit\Delta}\Pi_3$ by 
taking the finite size of the nucleus into account. While our 
subsequent treatment will not rely on any details of the spatial 
and temporal dependence of ${\mit\Delta}\Pi_3$, it is critical for a
reliable estimate of the magnitude of the 
kick induced by the chiral anomaly. Also, thus  estimate will play 
a role in our discussions on the signature of the anomaly effect 
in DCC, which will be given in Sec. V. 

For the purpose of taking account of the finiteness of the
nuclei, we consider a sphere with radius $R_A$ inside which
total charge $Z$ is uniformly distributed, as an abstract
of each colliding nucleus. We assume that two of such spheres are
colliding at Lorentz factor $\gamma$ and impact parameter $b$
in the center of mass frame,
where each nucleus appears to be a sphere Lorentz contracted with
the factor $\gamma$, or a pancake. We name the two pancake, pancake 1
and pancake 2.
In this case, $\vec{E}\cdot\vec{B}$ is given as follows:

1) if $\vec{r}$ is inside pancake 1, i.e., 
$R_1^2 \le R_A^2$,
and inside the pancake of nucleus 2, i.e.,
$R_2^2 \le R_A^2$,
\begin{equation}\label{finite1}
\vec{E}\!\cdot\!\vec{B} = \frac{2Z^2 e^2}{(4\pi)^2 M}\frac{\gamma^2}{R_A^6}
(\vec{r}\!\cdot\!\vec{L}),
\end{equation}
2) if $\vec{r}$ is inside pancake 1 and
outside pancake 2,
\begin{equation}\label{finite2}
\vec{E}\!\cdot\!\vec{B} =
\frac{2Z^2 e^2}{(4\pi)^2 M}\frac{\gamma^2}{R_A^3 R_2^3}
(\vec{r}\!\cdot\!\vec{L}),
\end{equation}
3) if $\vec{r}$ is outside pancake 1 and
inside pancake 2,
\begin{equation}\label{finite3}
\vec{E}\!\cdot\!\vec{B} =
\frac{2Z^2 e^2}{(4\pi)^2 M}\frac{\gamma^2}{R_1^3 R_A^3}
(\vec{r}\!\cdot\!\vec{L}),
\end{equation}
4) if $\vec{r}$ is outside pancake 1 and
outside pancake 2,
\begin{equation}\label{finite4}
\vec{E}\!\cdot\!\vec{B}
= \frac{2Z^2 e^2}{(4\pi)^2 M}\frac{\gamma^2}{R_1^3 R_2^3}
(\vec{r}\!\cdot\!\vec{L}) .
\end{equation}\\[2ex]
Using Eqs. (\ref{finite1}) - (\ref{finite4}), we obtain
${\mit\Delta}\Pi_3 \sim 0.03 m_{\pi}^{-2}$ in the central
region of a non-central collision at RHIC,
i.e., $z\sim 0$, $b\sim r_\perp \sim R_A = R({\rm Au})$,
$\vec{r} \perp \vec{b}$, and
$\gamma \sim 100$. The kick is substantially larger at LHC.

\section{Numerical Calculation --- Preliminaries}
\label{sec_num}

\subsection{Initial Condition and Constraints on Fields}
\label{sec_cal1}
We have carried out numerical calculations on a lattice of
$128\times 128$, on which we have imposed periodic boundary
conditions. The lattice is assumed to extend in the transverse
directions perpendicular to the collision axis.
In the longitudinal direction, the chiral
fields are assumed to be boost invariant as in \cite{ahw95}.
The lattice spacing is $a$. To include the initial correlation,
we have adopted an initial correlation length, 
$\ell_{\rm corr}$, which is generally larger than $a$.
The initial fields and
initial conjugate fields are set uniform within
$\ell_{\rm corr}\times\ell_{\rm corr}$ cells and
there is no inter-cell correlation.
We have also carried out simulations with the quench initial condition
obtained from physically thermalized field configurations \cite{ahw98}
and found that the result is insensitive to the details of the
initial condition.

The initial condition for the conjugate fields has been paid little 
attention to up to now. The initial conditions for both the chiral fields 
and the conjugate chiral fields, however, 
are expected to greatly affect the time evolution of 
the field configuration.  Thus, we first consider the problem of the 
initial condition for the conjugate fields.

We start with the following assumption on the chiral fields:
the chiral fields can be treated statistically even in the case
of the quench initial condition \cite{rw93,ahw95}.  This means that 
even if the initial field fluctuation is small, the fields have 
experienced some ergodic processes during the preceding phases of 
the heavy ion collision. 

The Hamiltonian density, ${\cal H}(\vec{x}, t)$, of the linear
sigma model above $T_c$, is given by
\begin{equation}\label{hamilton1}
{\cal H}(\vec{x}, t) =
\frac{1}{2}|\vec{\Pi} (\vec{x}, t)|^2 + 
\frac{1}{2}|\nabla \vec{\phi}(\vec{x}, t)|^2 + 
\frac{\lambda}{4}(|\vec{\phi}(\vec{x}, t)|^2
+w^2)^2- H\sigma,
\end{equation}
We have replaced $-v^2$ by $w^2$ in order to take account of
the chiral restoration in Eq. (\ref{hamilton1}).
In the mean field approximation,
$w^2$ and $v^2$ are related by \cite{ahw95,gm94},
\begin{equation}
w^2 = 3\langle\delta\phi^2_\parallel\rangle + \langle\delta\phi^2_\perp\rangle
-v^2,
\end{equation}
where $\delta\phi_{i\parallel}$ $(i=0 - 3)$
is the component of the fluctuation 
parallel to $\phi_i$ and $\delta\phi_{i\perp}$ is the orthogonal component.
In (\ref{hamilton1}), ${\vec{\Pi}}$ is the conjugate field of the
$\vec{\phi}$ field and should not be confused with the pion field.
Since we are assuming that initially the system is in the symmetric phase
and that the effective potential is convex, the motion of
the chiral fields is obviously bounded. Accordingly, the virial theorem
holds,
\begin{equation}\label{virial1}
\langle\Pi_i(\vec{x}, t)^2 \rangle_t
= \left \langle\phi_i (\vec{x}, t)
\frac{\partial {\cal H}(\vec{x}, t)}{\partial \phi_i(\vec{x}, t)}
\right\rangle_t ,
\end{equation}
where we have indicated that the average is a time average. Also note
that there is no implied summation over multiple indices $i$ in this 
subsection.
Equation (\ref{virial1}) leads to
\begin{equation}\label{virial2}
\langle\Pi_i(\vec{x}, t)^2 \rangle_t
= \langle\phi_i(\vec{x}, t)
(-\nabla^2 + \mu (\vec{x}, t)^2)\phi_i(\vec{x}, t)\rangle_t,
\end{equation}
with
\begin{equation}
\mu(\vec{x}, t)^2 = \lambda
(|\vec{\phi}(\vec{x}, t)|^2 + w^2).
\end{equation}
We have neglected a term proportional to $H$ in (\ref{virial2}). 

In numerical calculation, the Laplacian is discretized, and this affects
the virial theorem, (\ref{virial2}).  In both of the methods used in our 
numerical calculations we shall discuss in the next subsection, the second 
derivative in the spatial directions is discretized by
\begin{equation}
\nabla_i^2 f(\vec{x}) =
\frac{f(\vec{x}+a\vec{n}_i)+f(\vec{x}-a\vec{n}_i)-2f(\vec{x})}{a^2},
\end{equation}
where $\vec{n}_i$ is the unit vector in the $i$-th direction.  
For the correlation length $\ell_{\rm corr}= 2a$, we find
\begin{equation}\label{virial3}
\langle\phi_i(\vec{x}, t)\nabla^2\phi_i(\vec{x}, t)\rangle_t
= -\frac{D}{a^2}
\langle\phi_i(\vec{x}, t)^2\rangle_t,
\end{equation}
where $D$ is the spatial dimension, and the virial theorem becomes
\begin{eqnarray}
\langle\Pi_i(\vec{x}, t)^2\rangle_t
&= &\left( \frac{D}{a^2} + \mu^2 \right )
\langle\phi_i(\vec{x}, t)^2\rangle_t \nonumber\\
&\approx & \frac{D}{a^2}
\langle\phi_i(\vec{x}, t)^2\rangle_t. \label{virial4}
\end{eqnarray}
We assume that the field fluctuations at the initial time also
satisfy the spatial analog of (\ref{virial4}),
\begin{equation}
\langle\Pi_i(\vec{x}, t)^2\rangle_{\rm cell}
\approx \frac{D}{a^2}
\langle\phi_i(\vec{x}, t)^2\rangle_{\rm cell}, \label{virial5}
\end{equation}
where the average is taken over uncorrelated cells and events.  
We use the value $\ell_{\rm corr} = 2a$ in our numerical simulations 
throughout this paper. 

\subsection{Calculational Procedure}
\label{sec_cal2}
We have solved the equations of motion with the initial condition
discussed in the previous subsection. In numerical calculation,
the equations of motion are discretized. It is well-known that
naively discretized partial differential equations can be numerically
highly unstable, even if the underlying original continuum version of 
the equations is stable.  On the other hand, what is essential in the 
time evolution of DCC domains is the amplification of low momentum modes, 
or in other words, the physical instability.  
Thus, we have to satisfy the following two 
apparently contradictory requirements in solving the equations of 
motion on discretized lattice at the same time: eliminate unphysical 
instabilities, but retain the physical ones. In order to achieve this
goal, algorithms such as the Lax method and leap-frog method are often 
adopted.  However, such algorithms introduce quite often numerical 
viscosity. 
As a result, unwanted spurious suppression of fluctuations of short 
wave lengths occurs.  This does not cause a practical problem provided 
that the lattice spacing is small enough compared to the typical scale 
of the spatial variation of the fields.  In the simulation of the formation 
of DCC domains, however, the behavior matters because the fluctuations 
of short wave lengths are responsible for the change in the effective 
potential \cite{ahw95} and they affect the phase transition. 
Therefore, it is crucially important to keep short-range fluctuations 
to discuss the time evolution of DCC domains. 

In order to satisfy these requirements, we have used the following 
two methods, 
(i) the first order Adams-Bashforth method \cite{numrec} and 
(ii) the staggered leapfrog method for the second order term and 
the Crank-Nicholson method for the first order term in time\cite{press89}.  
Since (ii) is described in \cite{press89}, we briefly explain (i) here. 
Detailed discussions on the numerical calculations will be presented 
elsewhere \cite{ahw98}.  In both methods, if the lattice spacing is 
the same for all the spatial directions, 
the Laplacian is discretized as
\begin{eqnarray}
\triangle f (x_i,y_j,\tau_n)
& \rightarrow &
(f(x_{i+1},y_{j},\tau_n) + f(x_{i-1},y_{j},\tau_n) \nonumber\\
& & \quad + f(x_{i},y_{j+1},\tau_n) + f(x_{i},y_{j-1},\tau_n)
-4f(x_{i},y_{j},\tau_n))/a^2,
\end{eqnarray}
where $\tau$ is the proper time defined by $\sqrt{t^2 - z^2}$.
Note that only the
two dimensional Laplacian appears in the equations of motion
since we have assumed the longitudinal boost invariance
\cite{ahw95,bjorken2}.
The difference is in how
to carry out the time integration. In the Adams-Bashforth method,
the increment of $f$, $f(x_{i},y_{j},\tau_{n+1})-f(x_{i},y_{j},\tau_{n})$,
is given by \cite{dah74}
\begin{eqnarray}
f(x_{i},y_{j},\tau_{n+1})-f(x_{i},y_{j},\tau_{n})
& = & {\mit\Delta}\tau ( \dot{f}(x_{i},y_{j},\tau_{n})
+ \frac{1}{2}\bigtriangledown\! \dot{f}(x_{i},y_{j},\tau_{n})
+ \frac{5}{12}\bigtriangledown^2 \! \dot{f}(x_{i},y_{j},\tau_{n})\nonumber \\
&  & \quad\quad
        + \frac{3}{8}\bigtriangledown^3 \! \dot{f}(x_{i},y_{j},\tau_{n})
        + \frac{251}{720}\bigtriangledown^4 \! \dot{f}(x_{i},y_{j},\tau_{n})
        + \cdots ), \label{adba}
\end{eqnarray}
where ${\mit\Delta}\tau$ is the increment in $\tau$,
$\dot{f}(x_{i},y_{j},\tau_{n})$ is the proper time derivative of
$f(x_{i},y_{j},\tau_{n})$, and $\bigtriangledown^k$ is the
backward difference operator defined by
\begin{eqnarray}
\bigtriangledown g(x_{i},y_{j},\tau_{n})
& = & g(x_{i},y_{j},\tau_{n}) - g(x_{i},y_{j},\tau_{n-1}), \nonumber \\
\bigtriangledown^{k+1} g(x_{i},y_{j},\tau_{n})
& = & \bigtriangledown (
\bigtriangledown^k g(x_{i},y_{j},\tau_{n}) -
\bigtriangledown^k g(x_{i},y_{j},\tau_{n-1})).
\end{eqnarray}
The time integrals of $\sigma(x_i,y_j,\tau_n)$,
$\dot{\sigma}(x_i,y_j,\tau_n)$,
$\vec{\pi}(x_i,y_j,\tau_n)$,
and $\vec{\dot{\pi}}(x_i,y_j,\tau_n)$ are calculated
at each lattice site with the help of (\ref{adba}). 
In order to achieve satisfactory suppression 
of the unphysical instability, it is necessary to use sufficiently small 
${\mit\Delta}\tau$. For the typical parameters for the lattice spacing, 
the initial field fluctuation, and so on, which we shall specify shortly, 
we have found that ${\mit\Delta}\tau$ needs to be at least 1/1800 fm
or smaller. We have found that
the first order Adams-Bashforth method, which uses the first two terms
in the series on the RHS of Eq. (\ref{adba}), is good enough to suppress
the unphysical instability. 
In the following calculations, we adopt the above value for 
${\mit\Delta}\tau$.
We have also found that the required fineness of ${\mit\Delta}\tau$
in the method (ii) is similar and have verified that the two
methods give similar results.

\section{Numerical Calculation --- Results and Discussions}
\label{sec_num2}
\subsection{Parameters}
\label{sec_cal3}
In the following calculations, we shall use $\lambda=19.97$,
$v=87.4$ MeV, and $H=(119~{\rm MeV})^3$. These values correspond
to the pion mass $m_\pi = 135$ MeV, the sigma mass $m_\sigma=600$ MeV,
and the pion decay constant $f_\pi = 92.5$ MeV at $T=0$. We have
assumed the longitudinal boost invariance \cite{bjorken2}
at the initial time $\tau_0$, which is fixed at 1 fm, and we have fixed 
the lattice spacing $a = 0.25$ fm throughout the calculations.
The initial $\phi^i$ and $\dot{\phi}^i$ \cite{footnote1}
fields in each correlated cell are randomly distributed
according to a Gaussian form. In order to take account of
the finiteness of the initial hot system in the transverse
directions, we have adopted the following Gaussian noise parameters:
\begin{eqnarray}
\langle\sigma\rangle &=& (1-f(r))f_{\pi}, \nonumber\\
\langle\pi_i\rangle  &=& 0, \nonumber \\
\langle\sigma^2\rangle - \langle\sigma\rangle^2
&=& \langle\pi_i^2 \rangle = \delta_0^2 f(r), \nonumber \\
\langle\dot{\sigma}\rangle &=& 
\langle\dot{\pi}_1\rangle = \langle\dot{\pi}_2\rangle = 0, \nonumber\\
\langle\dot{\pi}_3\rangle &=& {\rm sgn}(y)a_n m_\pi^2 f(r), \nonumber \\
\langle\dot{\sigma}^2 \rangle&= &\langle\dot{\pi}_i^2 \rangle =
\frac{D\delta_0^2}{a^2}f(r), \label{initial1}
\end{eqnarray}
where $r$ is the distance from the origin to the center of a correlated 
cell and $\delta_0^2$ is a constant. In the following calculations, we use 
$\delta_0^2 = v^2 /16$, which corresponds to the quench scenario defined 
in \cite{ahw95}.
In relating $\langle\dot{\sigma}^2 \rangle$
and $\langle\dot{\pi_i}^2 \rangle$ to $\langle{\sigma}^2 \rangle$
and $\langle{\pi_i}^2 \rangle$, we have taken advantage of
the virial theorem (\ref{virial5}). $f(r)$ is an interpolation
function defined by
\begin{equation}
f(r) = \left [ \exp\left( \frac{r-R_0}{\Gamma}\right ) +1 \right ]^{-1}
\end{equation}
and describes the boundary condition. $R_0$ is the radius of the
initially excited region where fluctuations of the classical
chiral fields exist and the mean fields are different from
their values in the vacuum. Outside this region,
the chiral fields take the vacuum configuration,
$(f_\pi, \vec{0})$. $\Gamma$ is the thickness of the transient
region. The results presented in this paper are obtained
with $\Gamma = 0.5 $ fm. $a_n$ has been introduced to take account
of the effect of the anomaly kick discussed in Section \ref{sec_kick}.
$a_n$ is, in principle, dependent on the transverse coordinate.
We, however, assume that it is a constant in order to concentrate
on the effect of the kick to the time evolution of the system.
Thus, the initial $\dot{\pi}_3$ field consists of two parts,
a randomly fluctuating part and a part with global coherence. 
In Eq. (\ref{initial1}), we have defined $y$ direction to be
perpendicular to the scattering plane.
Following the estimate in Sec. II,
we shall take $a_n = 0.1$ in the calculation,  
which is of similar order but a bit larger than the one
expected at RHIC, but much smaller than the one
expected at LHC.
We note that this value is quite a moderate one as we see in the 
next subsection.

\subsection{Results I --- Effect of the Kick}
\label{sec_cal4}
In Figs. 1 and 2, we show an example of initial $\dot{\pi}_3$ and
$\dot{\pi}_2$ field configurations, respectively. The $x$ axis 
is parallel to the scattering plane, and perpendicular 
to the collision axis which is taken along the $z$ direction. 
In this and the following calculations, we assume that the system
extends infinitely in the transverse directions, i.e., $R_0 =\infty$,
unless otherwise specified. With $a_n = 0.1$ it is almost impossible 
to recognize the embedded coherent kick, $\langle\dot{\pi}_3\rangle = 
{\rm sgn}(y)a_n m_\pi^2$ in Eq. (\ref{initial1}), 
in the configuration of the $\dot{\pi}_3$ field.  
In Fig. 3, we show another initial configuration of 
$\dot{\pi}_3$, where the value of the kick has 
been artificially increased by a factor of five to $a_n$ = 0.5. 
It is only for illustrative purpose; if the value of $a_n$ were that 
large, it would be possible to distinguish the initial configurations
of $\dot{\pi}_2$ and $\dot{\pi}_3$. However, such a large value of $a_n$ 
is not what is suggested by the real world
at least at RHIC,
and the value we use in our simulations is $a_n$ = 0.1.  

At a glance of Figs. 1 and 2, it appears that the effect of the 
anomaly kick may be too tiny to be observable. However, two of us 
argued in Ref. \cite{mm96} that it might produce non-negligible 
effects on the evolution of the system because it is coherent 
over the nuclear dimensions. Moreover, the equations of motion for 
the chiral fields are coupled and highly non-linear, and therefore 
developing the physical intuition out of them would be quite difficult. 
Thus, we have decided to simulate the system numerically to uncover 
the time evolution of the $\pi_2$ and $\pi_3$ fields with the initial 
conjugate field configurations with the effect of the anomaly kick. 
It appears to the authors that it is the only tractable way of
answering to the question of how the anomaly kick is effective
in triggering DCC domains. 

The result we obtained is quite an unexpected one; 
the effect turns out to be significant. 
We show the proper time evolution of $\pi_3$ and $\pi_2$, respectively,
in Figs. 4 and 5 every 3 fm from $\tau_0$ in an event.
Fig. 4 indicates that the small kick results in very prominent
asymmetry of the $\pi_3$ field between the upper and lower
half-spaces. On the other hand, as is shown in Fig. 5,
the $\pi_2$ field does not show any upper-lower space asymmetry
in the course of the time evolution.
We have also carried out calculation
with $a_n = 0$ and have found that the time evolution of the
$\pi_2$ field is not affected by changing the value of $a_n$
\cite{footnote2}. 
Thus, the motion of different isospin components of pion fields 
effectively decouples from each other, and the anomaly kick affects 
only the time evolution of $\pi_3$. We will explain the reason for
this effective decoupling below.

In addition to the existence of the upper-lower asymmetry, we find that 
the coherent structure which produces this asymmetry oscillates in time.
The medium scale structure which both figures have in common corresponds 
to what is usually referred to as the DCC.  Figs. 4 and 5 tell us that 
the anomaly kick induces a spatial structure in the $\pi_3$ field 
configuration which is larger than the ordinary DCC domains. This can 
also be observed in Figs. 6 and 7, in which we have plotted the Fourier 
power of $\pi_3$ and $\pi_2$ at $\tau = \tau_0 + 3 ~{\rm fm} = 
4 ~{\rm fm}$, respectively.
The scale is arbitrary, but the same scale has been used for
Figs. 6 and 7. Note that this value of $\tau$ corresponds 
to the second proper time in Figs. 4 and 5. Both figures show a strong 
enhancement of low momentum modes, but in Fig. 6 we observe further sharp 
enhancement at very low momenta. The former is due to the amplification of
low momentum modes in the formation of normal DCC domains first discussed 
in Ref. \cite{rw93}, and the latter is due to the asymmetric coherent 
collective oscillation of the $\pi_3$ field induced by the anomaly kick.

How can we understand such large effect in the time evolution of
the $\pi_3$ field caused by such a small kick?
The reason is twofold. 
First, it is because the different field modes are effectively
decoupled in the quench case. The equations of motion for the
fields are non-linear, but the interaction term in the action is 
proportional to $(\phi^i)^2(\phi^j)^2$. 
Thus, it is expected that in the quench case, where the fluctuation 
of the fields is small, the approach toward the equipartition 
is slow. This is a simple analog of the fact that the relaxation 
time is longer at lower temperatures in the kinetic theory.  
Accordingly, low momentum modes behave as if they were almost 
independent oscillators undisturbed by the rest of the system. 
Also, in the quench scenario, the fields are initially concentrated 
at the local maximum of the effective potential with little 
fluctuation, and so even a small kick is expected to be effective 
in determining the subsequent motion of the fields if it is coherent 
in space.

The second reason is the weak coupling of Nambu-Goldstone modes
at low momenta. When there exists expansion, the field 
configurations approach to the chiral circle quickly and
gets concentrated around the chiral circle
after a very short time \cite{rw93,ahw95}. As is well-known,
if the chiral fields are constrained on the chiral circle, 
the effective coupling among the pion fields becomes of the form
of $\sim(\vec{\pi}\!\cdot\!\partial_\mu \vec{\pi})^2$. As a result,
modes with small momenta are isolated from the rest of the system,
and the effect of the coherent kick is not much affected by the presence
of fluctuations of large momenta. 

\subsection{Results II --- Currents}
\label{sec_cal5}
In the previous subsection, we have observed that the anomaly kick
greatly affects the time evolution of the chiral fields. However,
the field strengths themselves are not physical observables. 
The physical observables are currents, particle numbers, particle
distributions, and so forth. In this subsection, we consider
the behavior of physical currents.

The $O(4)$ sigma model possesses an $SU(2)\times SU(2)$ symmetry
in the limit of $H\rightarrow 0$. Correspondingly, two current densities,
the vector current density $V_{\mu}^{i}(\vec{x},t)$ and axial current
density $A_{\mu}^{i}(\vec{x},t)$, where $i$ and $\mu$ are isospin
and Lorentz indices, respectively, 
are defined as
\cite{gl,iz80}:
\begin{eqnarray}
V_{\mu}^{i}(\vec{x},t) & = & \varepsilon^{ijk}\pi^j (\vec{x},t)
\partial_{\mu}\pi^k (\vec{x},t), \nonumber \\
A_{\mu}^{i}(\vec{x},t) & = &
\pi^i (\vec{x},t)\partial_{\mu}\sigma(\vec{x},t)
- \sigma(\vec{x},t)\partial_{\mu}\pi^i (\vec{x},t). \label{currents1}
\end{eqnarray}
They satisfy the CVC and PCAC, respectively,
\begin{eqnarray}
\partial^\mu V_{\mu}^i (\vec{x},t) & = & 0, \nonumber \\
\partial^\mu A_{\mu}^i (\vec{x},t) & = & H \pi^i (\vec{x},t).
\label{currents2}
\end{eqnarray}
In the following, we calculate averaged charge densities,
i.e., the average of the zeroth component of the current densities.
The average is taken separately over the upper and lower half-spaces 
with respect to the scattering plane. 
Average over events is taken also to reduce fluctuations.
Since the anomaly kick affects only the $\dot{\pi}_3$ field, 
its effect is expected to appear only in the averages of 
$V_{0}^{1}$, $V_{0}^{2}$, and $A_{0}^{3}$. Other charges
are not expected to show any upper-lower asymmetries. 

In Fig. 8, we show $\langle A_{0}^{1}\rangle_{\rm upper}$,
$\langle A_{0}^{3}\rangle_{\rm upper}$, and 
$\langle A_{0}^{3}\rangle_{\rm lower}$,
as a function of the proper time. 
The average is taken over 10 events in Figs. 8 - 10. 
We observe that $A_{0}^{3}$ shows distinct upper-lower
asymmetry. On the other hand, $A_{0}^{1}$ does not show it.
This, however, does not mean that $A_{0}^{1}$ does not show
domain structure. As is shown in Ref. \cite{ammr98},
the low momentum modes of all the axial charges get amplified without the
anomaly kick in the course of the time evolution from the quench
initial condition. The reason for the lack of upper-lower asymmetry in 
the variable $A_0^1$ is the absence of such an asymmetry in the initial 
field configuration.  In Fig. 9, $\langle V_{0}^{1}\rangle_{\rm upper}$ and
$\langle V_{0}^{3}\rangle_{\rm upper}$ are shown. The vector charges do not 
show any asymmetry.  Since the sign of the kick is fixed in our
calculation, the expectation values of $A_{0}^{3}$ in the upper and lower 
half-spaces do not vanish after taking average over events.

The behaviors of $\langle A_{0}^{1}\rangle_{\rm upper}$,
$\langle A_{0}^{3}\rangle_{\rm upper}$,
and $\langle A_{0}^{3}\rangle_{\rm lower}$
are better understood by comparing them with the behaviors of
$\langle \pi_1\rangle_{\rm upper}$, 
$\langle \pi_3\rangle_{\rm upper}$,
and $\langle \pi_3\rangle_{\rm lower}$,
which are shown in Fig. 10.  
By comparing Fig. 8 and Fig. 10, we find that the extrema of 
$\langle A_{0}^{3}\rangle_{\rm upper}$ and 
$\langle A_{0}^{3}\rangle_{\rm lower}$ approximately correspond to the 
zero points of $\langle \pi_3\rangle_{\rm upper}$ and 
$\langle \pi_3\rangle_{\rm lower}$.  
This is what is expected from PCAC, which says that 
\begin{equation}
\partial^0 \langle A_{0}^3 (\vec{x},t) \rangle = 
H \langle \pi^3 (\vec{x},t) \rangle 
+ {\rm boundary~terms} .
\end{equation}
An alternative and consistent interpretation of the above feature 
is that the time evolutions of
$\langle \pi_3\rangle_{\rm upper}$ and
$\langle \pi_3\rangle_{\rm lower}$
are approximated by harmonic oscillations as seen from Fig. 10, 
and thus at the zero points
$\langle \dot{\pi}_3\rangle_{\rm upper}$ and
$\langle \dot{\pi}_3\rangle_{\rm lower}$ take extreme values. On the 
other hand, the sigma field does not receive any kick and, accordingly, 
does not show upper-lower asymmetry. Also, the oscillation of the 
sigma field is quickly damped and the sigma field approaches 
$f_\pi$\cite{ahw95,hw94}.  From this observation and the definition of the 
axial currents, Eq. (\ref{currents1}), it is thus concluded that the 
collective oscillation of the $\pi_3$ field induced by the kick is 
responsible for the upper-lower asymmetry and the oscillatory behavior of
$\langle A_{0}^{3}\rangle$.

The non-vanishing, upper-lower asymmetric, and oscillatory
feature of $\langle A_{0}^{3}\rangle$
survives even if the transverse extent of the system is
finite. In Fig. 11, we show  
$\langle A_{0}^{1}\rangle_{\rm upper}$,
$\langle A_{0}^{3}\rangle_{\rm upper}$,
and $\langle A_{0}^{3}\rangle_{\rm lower}$,
for the case with $R_0 =5$ fm. The average is taken
over all grid points with $r \le 5 $ fm and over 100 events.
In this case, the amplitudes of the oscillations of 
$\langle A_{0}^{3}\rangle_{\rm upper}$ and
$\langle A_{0}^{3}\rangle_{\rm lower}$ are
smaller than in the $R_0 = \infty$ case.
This is due to the reduction of the strengths of the pion
fields caused by the transverse expansion.

\subsection{Results III --- Correlation Functions}
\label{sec_cal6}

In section \ref{sec_cal4}, we studied the time evolution of the chiral 
fields and their coherence in coordinate space visually by plotting the 
field strengths of the chiral fields. The best way of quantifying the
coherence of fields is to calculate correlation functions.  In this 
subsection, we study the correlation functions of the chiral fields and 
charges.

First, we define the correlation function $A_{ij}(\vec{r}, t)$ by
\begin{equation}
\label{corr1}
A_{ij}(\vec{r},t) = \frac{1}{V}\!\int\!
\pi_i(\vec{x}, t)\pi_j(\vec{x}+\vec{r}, t)d^3 x
-\frac{1}{V^2}\!\int\!\pi_i(\vec{x},t)d^3 x
\!\int\!\pi_j(\vec{x}',t)d^3 x',
\end{equation}
where $V$ is the volume of the system.  Since the fields $\pi_i$ form the
basis of a triplet representation of $SU(2)$, $A_{ij}(\vec{r},t)$ can be 
decomposed as \cite{sakurai}
\begin{equation}
A_{ij}(\vec{r},t) = \delta_{ij}{\cal S}(\vec{r},t) + 
\varepsilon_{ijk}{\cal V}_k(\vec{r},t) + {\cal T}_{ij}(\vec{r},t),
\end{equation}
where ${\cal S}(\vec{r},t)$, ${\cal V}_k(\vec{r},t)$, and 
${\cal T}_{ij}(\vec{r},t)$ are given by
\begin{eqnarray}
{\cal S}(\vec{r},t) & = &
\frac{1}{3}A_{kk}(\vec{r},t),\nonumber \\
{\cal V}_k(\vec{r},t) & = &
\frac{1}{2}\varepsilon_{ijk}A_{ij}(\vec{r},t), \nonumber \\
{\cal T}_{ij}(\vec{r},t) & = &
\frac{1}{2}(A_{ij}(\vec{r},t) + A_{ji}(\vec{r},t))
- \delta_{ij}{\cal S}(\vec{r},t).
\end{eqnarray}
If $A_{ij}(\vec{r},t) = A_{ji}(-\vec{r},t)$ holds,
${\cal V}_k(\vec{r},t)$ vanishes.

Next, let us assume a solution of the equation of motion for the
$\pi_i$ field, $\pi_i(\vec{x},t)$, with given initial condition at
$t=t_0$.  Suppose the sign of the initial conditions for the $\phi_i$ and 
$\dot{\phi}_i$ fields is reversed, $\phi_i(\vec{x}, t_0) \rightarrow - 
\phi_i(\vec{x},t_0)$, $\dot{\phi}_i(\vec{x}, t_0) \rightarrow - 
\dot{\phi}_i(\vec{x},t_0)$.  Then, the solution for the $\pi_i$ field in 
this case is just $-\pi_i(\vec{x},t)$. This is because the interaction term 
in the equation of motion for $\pi_i$ is proportional to $(\sigma^2 + 
\pi_1^2 + \pi_2^2 + \pi_3^2 )\pi_i$.  Thus, if the probability
distribution $P(\pi_k(\vec x), \dot\pi_k(\vec x))$ of the initial values
of the fields $\pi_k$ and $\dot\pi_k$ satisfies
\begin{equation}
P(\pi_k(\vec x),\dot\pi_k(\vec x)) = P(-\pi_k(\vec{x}), 
-\dot{\pi}_k(\vec{x})), \label{corr2}
\end{equation}
the event averages of all the off-diagonal correlation functions 
$\langle A_{ij}(\vec{r},t)\rangle$ vanish, and hence 
$\langle {\cal T}_{ij}(\vec{r},t)\rangle = 0$ for $i\neq j$.  
In the case we are dealing with, only the $\dot{\pi}_3$ 
field is kicked, and for the $\pi_1$ and $\pi_2$ fields, 
the relationship (\ref{corr2}) still holds. Accordingly,
$\langle {\cal T}_{ij}(\vec{r},t)\rangle = 0$ for $i\neq j$.  
We have indeed confirmed this in our numerical calculation.
Hence, in the following, we shall only consider the diagonal 
correlation functions.

First, in Fig. 12, we show $A_{11}(r,\tau)$ at three different proper times.
We have used the default parameters and the result has been obtained
by averaging over 5 events. The average over the azimuthal angle of
$\vec r$ has been performed. We observe that 
the correlation function changes rather substantially as a function 
of the proper time. 
The pion fields oscillate and both the coherent part and
short distance fluctuation change in time. Due to the energy conservation, 
when the amplitude of the coherent motion is at the maximum, the local 
fluctuation is at the minimum.  $\tau=5$ fm approximately corresponds 
to this time and in a two dimensional plot domain structure is clearly 
observed without fluctuations of much shorter scale. 

One interesting feature of $A_{11}$ is the oscillatory behavior. This 
was already observed in Ref. \cite{ahw95}. It has been sometimes 
interpreted as a sign of the shrinkage of the domains or occurrence of 
anti-correlation.  However, this is not the correct interpretation. The
behavior is actually due to the pseudo-periodicity in the distribution of 
the pion field strengths. This can be seen, for instance, in Fig. 5.
The peaks of the pion field strength are not distributed randomly.
The reason for this is traced back to the mechanism responsible for the
formation of DCC domains, i.e., the amplification of low momentum pion modes.
The wavelengths of the low momentum modes characterize not only the size 
of each domain but also the distribution or separation of domains. In the 
mean field theory, modes with momentum less than a certain cutoff value are
amplified at each time and not only specific modes are amplified.
This is the reason why the distribution of the peaks is not utterly regular.
We note that a similar oscillatory behavior in correlation functions 
of systems without regular lattice structure (such as liquids) 
is also known in condensed matter physics \cite{Good75}.
DCC formation is often compared to the spinodal decomposition
observed in glass and metal \cite{cahn64,spinodal1}. It is known, however, 
that in the case of the spinodal decomposition only modes in a small range 
of momenta are selectively amplified with large amplification factors. As a 
result, the distribution of domains created by the spinodal decomposition 
generally shows more periodic structure. 

Next, we show $A_{33}(r,\tau)$ at $\tau=7$ and 9 fm in Fig. 13.
As we have seen that the time evolution of the $\pi_3$ field is
asymmetric between the upper and lower half-spaces, we distinguish
the same side and different side correlation functions, 
$A_{33}^{++}(r,\tau)$ and $A_{33}^{+-}(r,\tau)$.  
They are defined as follows:
\begin{eqnarray}
A_{33}^{++}(r,\tau) & = &
\langle(\pi_3(i,\tau) - \langle\pi_3(\tau)\rangle )
(\pi_3(j,\tau) - \langle\pi_3(\tau)\rangle )\rangle_{ij}
\quad {\rm sgn}\,y(i)\,{\rm sgn}\,y(j)>0, \nonumber\\
A_{33}^{+-}(r,\tau) & = &
\langle(\pi_3(i,\tau) - \langle\pi_3(\tau)\rangle )
(\pi_3(j,\tau) - \langle\pi_3(\tau)\rangle )\rangle_{ij}
\quad {\rm sgn}\,y(i)\,{\rm sgn}\,y(j)<0, \label{corr3}
\end{eqnarray}
where $i$ and $j$ are grid points, $\langle\cdots\rangle_{ij}$ is the 
average over pairs of those grid points, $i$ and $j$, such as the distance
between them is $r$ and the product of the signs of their $y$-coordinates 
is either positive $(++)$ or negative $(+-)$, and $\langle\pi_3(\tau)\rangle$
is the average of the strength of the $\pi_3$ field at $\tau$ over the 
whole space. We show the average of Eq. (\ref{corr3}) over 5 events in Fig. 
13. From Fig. 10, we see the amplitude of the collective motion of the 
$\pi_3$ field is about the maximum at $\tau = 7$ fm, and, accordingly, 
the upper-lower asymmetry is largest at that time.  Correspondingly, we 
can clearly see the asymmetry in the correlation function, $A_{33}$ at 
$\tau=7$ fm. On the contrary, at $\tau=9$ fm, there is little 
asymmetry observed in $A_{33}$. This is because, as is seen from Fig. 10, 
the amplitude of the coherent oscillation almost vanishes at this time. For 
comparison, we plot the same side and different side correlations of the 
$\pi_1$ field in Fig. 14 at $\tau = 7$ fm.  As in the previous figures, 
an average over 5 events has been taken. There is little difference 
between the two correlation functions. We have confirmed that this 
small amount of difference is solely due to statistical fluctuations 
by carrying out other sets of statistically independent simulations.

We have further calculated the correlation functions of the axial
and vector charges. We have defined the same side and different side
correlation functions of the axial charges, $B_{kl}(r,\tau)$, and
the vector charges, $C_{kl}(r,\tau)$, by
\begin{eqnarray}
B_{kl}(r,\tau)&=&
\langle(A_0^k (i,\tau) - \langle A_0^k (\tau)\rangle )
(A_0^l (j,\tau) - \langle A_0^l (\tau)\rangle )\rangle_{kl},\nonumber \\
C_{kl}(r,\tau)&=&
\langle(V_0^k (i,\tau) - \langle V_0^k (\tau)\rangle )
(V_0^l (j,\tau) - \langle V_0^l (\tau)\rangle )\rangle_{kl},
\end{eqnarray}
where the notations are the the same as before and
the average is taken over 5 events.
In Fig. 15, we show the same side and different side correlation
functions of $A_0^3$ at $\tau = 9$ fm.  We note that this proper 
time corresponds to when the asymmetry in $A_0^3$ is approximately at 
maximum as seen in Fig. 8. In Fig. 15, the asymmetry between the same 
side and different side correlation functions is observed. However, it 
should be noted that the degree of coherence is less than that of the 
field-field correlation functions.  In Fig. 16, same side correlation 
functions of $A_0^1$, $V_0^1$, and $V_0^3$ are plotted at $\tau=9$ fm.
This indicates that $A_0^1$ shows some coherence,
but that neither $V_0^1$ nor $V_0^3$
shows coherent behavior \cite{ammr98}.

\subsection{Results IV --- Pion Density Distributions}
\label{sec_cal7}
It has often been argued that the pion density in coordinate space is 
proportional to the square of the pion field strength.  This is actually 
the basis for the expression for the probability to have a DCC domain in 
which the fraction of neutral pions, ${n_{\pi^0}/(n_{\pi^0}+ n_{\pi^+}
+n_{\pi^-})}$, takes the value $R$ \cite{anselm,bjorken1,blaizot1},
\begin{equation}
P(R) = \frac{1}{2\sqrt{R}}.
\end{equation} 
In this formula, the pion fields are assumed to be static,
$\dot{\pi}_i=0$.  In reality, however, the pion fields are not static and 
also include substantial fluctuations. When the pion fields are not static, 
the pion densities depend also on the conjugate fields.  This is just an 
analog of the one dimensional harmonic oscillator case; both the coordinate 
variable and its canonical conjugate (momentum) variable contribute to the 
energy, i.e., the number of quanta.  Moreover, when the pion fields are 
fluctuating even classically, the definition of a domain itself is not 
unambiguous.  However, in the following we shall calculate the local pion 
densities in coordinate space as a function of 
proper time and discuss the typical 
size of clusters defined by the distribution of local pion densities 
thus obtained.

Quantum field theory is often formulated in momentum space.  Creation and 
annihilation operators are usually defined for momentum eigenstates. It is 
because the Hamiltonian for free fields reduces to the sum of those of 
harmonic oscillators in momentum space.  However, as we have discussed 
above, what we need here is the pion density distribution in coordinate 
space. For that purpose, we first define the pion creation operators and 
annihilation operators in coordinate space as the Fourier transforms of those
in momentum space. For example, the creation  operators for $\pi^+$, 
$\pi^-$, and $\pi^0$ at $\vec{x}$ are given, respectively, by
\begin{eqnarray}
a_{\pi^+}^{\dag} (\vec{x}) & = & \frac{1}{(2\pi)^{3/2}}
\!\int\! e^{-i\vec{p}\cdot\vec{x}}
\left( \sqrt{\frac{\omega_p}{2}}{\pi^+}^{\dag}(\vec{p})
-i\frac{1}{\sqrt{2\omega_p}}{\dot{\pi}}^{+\dag}(\vec{p})\right )
d^3 p, \nonumber \\
a_{\pi^-}^{\dag} (\vec{x}) & = & \frac{1}{(2\pi)^{3/2}}
\!\int\! e^{-i\vec{p}\cdot\vec{x}}
\left ( \sqrt{\frac{\omega_p}{2}}{\pi^-}^{\dag}(\vec{p})
-i\frac{1}{\sqrt{2\omega_p}}{\dot\pi}^{-\dag}(\vec{p})\right )
d^3 p, \nonumber \\
a_{\pi^0}^{\dag} (\vec{x}) & = & \frac{1}{(2\pi)^{3/2}}
\!\int\! e^{-i\vec{p}\cdot\vec{x}}
\left ( \sqrt{\frac{\omega_p}{2}}{\pi^0}^{\dag}(\vec{p})
-i\frac{1}{\sqrt{2\omega_p}}{\dot\pi}^{0\dag}(\vec{p})\right )
d^3 p, \label{annihi1}
\end{eqnarray}
where $\omega_p = \sqrt{m_\pi^2 + \vec{p}^2}$, $\pi_{i}(\vec{p})$ and 
${\dot\pi_i}(\vec{p})$ are, respectively, the Fourier transforms of
$\pi_{i}(\vec{x})$ and $\dot{\pi}_{i}(\vec{x})$. The charge eigenstate
fields $\pi^+(\vec{x})$, $\pi^-(\vec{x})$, and $\pi^0(\vec x)$ are defined as
\begin{eqnarray}
\pi^{\pm}(\vec{x}) &= &\frac{\pi_1(\vec{x})\pm i \pi_2(\vec{x})}{\sqrt{2}}, 
\nonumber \\
\pi^0(\vec{x}) & = & \pi_3(\vec{x}) . \label{pippim}
\end{eqnarray}
We define the local density operators for the $\pi_{i}$ field,
$n_{\pi_i}(\vec{x})$ as
\begin{equation}
n_{\pi_i}(\vec{x}) = a_{\pi_i}^{\dag}(\vec{x})a_{\pi_i}(\vec{x}).
\label{denscs1}
\end{equation}
We note that since
\begin{equation}
\pi^{\pm}(\vec{p}) \neq {\pi^\pm}^{\dag}(\vec{p}),
\end{equation}
in general,
\begin{equation}
n_{\pi^{+}}(\vec{x}) \neq  n_{\pi^-}(\vec{x}).
\end{equation}
Thus, in principle, the electric charge density can fluctuate and
can take non-zero values locally.
The creation operators and density
operators satisfy the following commutation relation:
\begin{equation}
[n_{\pi_i}(\vec{x}), a_{\pi_j}^{\dag}(\vec{x}')]
= \delta_{ij}\delta(\vec{x}-\vec{x}')a_{\pi_j}^{\dag}(\vec{x}').
\end{equation}
This justifies our definition of the creation, annihilation,
and density operators in coordinate space.
These creation operators do not create momentum or energy eigenstates,
since the position is specified. However, Parsival's relation
is satisfied,
\begin{equation}
\int\! n_{\pi_i}(\vec{x})d^3 x = \int\! n_{\pi_i}(\vec{p}) d^3 p,
\end{equation}
and, accordingly, the integrations over coordinate
space and momentum space give the same total number.

In this subsection, we have not yet specified the physical state 
of the system. In order to extract the coordinate
space distribution of each pion species and find the typical
size of DCC domains in the original sense, we first rewrite
$n_{\pi_i}(\vec{x})$ as follows:
\begin{equation}
n_{\pi_i}(\vec{x}) = \frac{1}{2}
\left |
\int \left(
P(\vec{x}-\vec{x}')\pi_{i}(\vec{x}')+
iQ(\vec{x}-\vec{x}')\dot{\pi}_{i}(\vec{x}')
\right )
d^3 x' \right |^2 ,
\label{denscs2}
\end{equation}
where $P(\vec{x})$ and $Q(\vec{x})$ are functions defined by
\begin{eqnarray}
P(\vec{x}) & = & \frac{1}{(2\pi)^3}\!\int\!\sqrt{\omega_p}
e^{i\vec{p}\cdot\vec{x}}d^3 p, \nonumber \\
Q(\vec{x}) & = & \frac{1}{(2\pi)^3}\!\int\!\frac{1}{\sqrt{\omega_p}}
e^{i\vec{p}\cdot\vec{x}}d^3 p.
\end{eqnarray}
The non-local feature of $n_{\pi_i}$ is obvious in 
Eq. (\ref{denscs2}).
We now assume that the system is in a coherent state
$|\,{c}\,\rangle$ and the operators $\pi(\vec{x})$ and
$\dot{\pi}(\vec{x})$ in $a(\vec{x})$ operating on $|\,{c}\,\rangle$
can be replaced by those values obtained in the classical simulation,
$\pi_{\rm cl}(\vec{x})$ and $\dot{\pi}_{\rm cl}(\vec{x})$.
Note that this procedure needs to be modified if the system is not in a 
coherent state \cite{kogan}. This possibility will be considered elsewhere.

In Figs. 17 and 18, we show the distributions of $\pi^0$ and $\pi^+$
density on the $x$-$y$ plane, respectively, every 3 fm from $\tau_0$
in an event. We have used the same initial field
configuration as for Figs. 4 and 5. From Figs. 17 and 18, we can
clearly see the formation of domains also in this definition.
However, we recognize in these figures several features different
from Figs. 4 and 5. First, the distributions of $\pi^0$ and $\pi^+$
densities fluctuate more than those of the $\pi^0$ and $\pi^+$ fields
themselves. This is because the definitions of $\pi^0$ and $\pi^+$
density include, respectively, $\dot{\pi}^0$ and $\dot{\pi}^+$ fields,
which are not so much amplified as $\pi^0$ and $\pi^+$ fields and
noisier than those. Second, the distribution of $\pi_0$ density
does not show the upper-lower asymmetry. This is because the definition 
of $n_{\pi_i}(\vec{x})$ is quadratic in $\pi_i(\vec{x})$ and 
$\dot{\pi}_i(\vec{x})$.  Third, after the initial formation of the large 
scale structure, the pattern of the distribution remains almost unchanged
except for the occurrence of slight diffusion. This tells us that
domains in this definition are almost frozen at fixed positions on the 
$x$-$y$ plane. The last point is also clearly seen in Fig. 19, where we have 
plotted the time evolution of the distribution of $R(\vec{x})=
n_{\pi^0}(\vec{x})/(n_{\pi^+}(\vec{x}) +n_{\pi^-}(\vec{x}) 
+n_{\pi^0}(\vec{x}))$ in the same event.

In Fig. 20, we have plotted $R$, which is defined as
\begin{equation}
R=\biggl.
{ \sum_{\rm event}\int n_{\pi^{0}}(\vec{k}) d^3 k }
\biggm/
{ \sum_{\rm event}\int
(n_{\pi^+}(\vec{k})+n_{\pi^-}(\vec{k})+n_{\pi^{0}}(\vec{k})) d^3 k }
\biggr. ,
\label{rave}
\end{equation}
where $n_{\pi_i}(\vec{k})$ is the pion density at momentum $\vec{k}$, in 
a solid line. Ten events were used in the average.  The dashed line is 
the same except that the momentum integration is limited to $k< 250$ MeV. 
Fig. 20 shows striking enhancement of neutral pions, especially in the 
low momentum region.  Note that $R=0.375$ corresponds to $N_{\pi^0} : 
N_{\pi^+} : N_{\pi^-} = 1.2 : 1 : 1$ provided $N_{\pi^+} =  N_{\pi^-}$.
Also note that the distribution of $R$ deviates from the celebrated
form, $P(R)=1/2\sqrt{R}$, when $\pi_i(\vec{x})$ is not uniform,
and moreover that even if $\pi_i(\vec{x})$ is spatially
uniform, when $\dot{\pi_i}(\vec{x}) \neq 0$, $R$ also depends on
$\dot{\pi_i}(\vec{x})$. 
Fig. 21 is the same as Fig. 20 except that no anomaly kick
is exerted. In this case, $R$ is, as is expected,
approximately 1/3 regardless of the cutoff in the momentum integral.

We have found that
the distribution of the excess $\pi^0$'s is not isotropic in momentum
space. They are distributed more in the $y$-direction than the
$x$-direction. This reflects the Fourier power of the initial
anomaly kick: in the $R_0 \rightarrow \infty$ limit, the Fourier
power of the kick, ${\rm sgn}(y)a_n m_\pi^2$, is concentrated
at $p_x = 0$ in the $x$-direction, while it
has support at $p_y \neq 0$ in the $y$-direction.
In addition, the effect of the anomaly kick is expected to be
observed more in the central rapidity region than in the
fragmentation region, because the kick is created
when the two colliding nuclei are almost overlapping with each other.

\section{Summary}
\label{sec_Sum}

In summary, we have studied the physical influence of the axial
anomaly on the formation mechanism of disoriented chiral
condensate domains in relativistic heavy ion collisions.
Our study is based on the
framework of the linear sigma model with initial conditions
corresponding to a ``quench'' modified by a small, spatially
coherent displacement (``anomaly kick'') of the conjugate field
of the neutral pion field caused by the action of the electromagnetic
fields of the colliding nuclei.
These initial conditions correspond to a somewhat schematic, but
quantitatively realistic abstraction of conditions that could arise
in semiperipheral collisions of two heavy nuclei at RHIC or LHC.

We have found that the anomaly kick has practically no influence on the 
evolution of the charged pion fields but causes noticeable modifications 
to the dynamics of the neutral pion field. It produces a spatial asymmetry
in the formation of DCC domains above and below the scattering plane
and leads to a general enhancement of the $\pi^0$ domain structure.
The up-down asymmetry is clearly expressed in the integrated $I_3=0$
component of the axial charge on both sides of the scattering plane.
Since the signs of the pion fields or the axial charges are not
directly connected with hadronic observables, this up-down asymmetry
is difficult to establish experimentally. 

However, the overall enhancement of domain formation in the $I_3=0$ 
direction is found to lead to a relative increase in the neutral pion 
yield, raising the ratio $R = N_{\pi^0}/N_{\pi}$ significantly above 
the value 1/3 dictated by isospin symmetry. This signature, which
survives event averaging, should be experimentally detectable.
Because it has a different characteristic than the isospin breaking
due to the effects caused by charges of incident nuclei, which alters 
the $\pi^+/\pi^-$ ratio,
it can be discriminated from this well known effect. Also, the mechanism 
of isospin breaking discussed here can be distinguished from
the pion mass splitting, which leads to a similar but smaller 
enhancement of $R$, because it depends on the impact parameter 
and vanishes for precisely central collisions.

Finally, we have analyzed spatial correlation functions of the pion
field. The oscillations already observed by others appear to be
caused by long-range correlations in the positions of different
DCC domains. These correlations correspond to a fluid-like spatial
structure of the domain locations and are clearly visible in equal-time
snapshots of the chiral order parameter.

If the domain formation of DCC is in fact triggered by the anomaly 
effect discussed in this paper, there is a clear experimental 
signature which testifies for its origin; the formation rate of DCC 
should show a strong energy and atomic number dependences 
characteristic to the mechanism. The parameter $a_n$ which measures 
the magnitude of the anomaly kick scales as $E$ and $Z^2$, respectively, as 
center of mass energy $E$ and the atomic number $Z$ are varied. 
It is tempting to speculate that the formation rate of DCC somehow 
should reflect these dependences.  

Experimental searches for DCC formation have, so far, concentrated
on central collisions between two heavy nuclei, because it is here
where one expects the highest energy densities to occur over the
largest spatial volumes. These two conditions are important, because
matter needs to be heated above $T_c$ over a large volume if one
wants to observe the characteristic effects of a phase transition.

However, our results indicate that semi-peripheral collisions,
i.e., collisions with an impact parameter $b\sim R_A$, may be even 
more favorable for DCC formation. Two reasons
contribute to this phenomenon: (1) The expansion time scale is 
generally shorter in peripheral collisions than in central ones, 
and a three-dimensional expansion may occur more easily in non-central 
collisions, leading to the quench initial condition.
(2) Collisions of heavy nuclei with $b \neq 0$ produce 
the ``anomaly kick'', which favors the formation of large uniform domains 
of the quark condensate by inducing a collective motion in the
direction of the neutral pion field. Therefore, we suggest that
the search for DCC formation in relativistic heavy ion collisions
at RHIC should not be confined to central collisions only.

\acknowledgments

We thank Krishna Rajagopal for many valuable communications and 
for sharing with us useful knowledges through the collaboration 
on the work \cite{ammr98}.
M. A. and H. M. would like to thank for warm hospitality at 
Duke University, and M. A. and B. M. extend their gratitude to 
the members of Physics Department at Tokyo Metropolitan University 
for warm hospitality and fruitful discussions. M. A. wishes
to thank the Institute for Nuclear Theory at the University of Washington
for its hospitality during his stay.
This research has been supported in part by the U.S. Department of
Energy under Grant DE-FG02-96ER40945.
H. M. is partly supported by Grant-in-Aid for Scientific Research
no. 09640370 of the Japanese Ministry of Education, Science and Culture,
and by Grant-in-Aid for Scientific Research no. 09045036 under the
International Scientific Research Program, Inter-University
Cooperative Research.


\newpage

\centerline{\small\bf FIGURE CAPTIONS}
\vspace*{3ex}

\noindent
Fig. 1: Example of initial $\dot{\pi}_3$ field configuration.

\noindent
Fig. 2: Example of initial $\dot{\pi}_2$ field configuration.

\noindent
Fig. 3: Example of initial $\dot{\pi}_3$ field configuration. The value
of the kick has been artificially increased by a factor of five
to $a_n = 0.5$.

\noindent
Fig. 4: Proper time evolution of the $\pi_3$ field in an event.

\noindent
Fig. 5: Proper time evolution of the $\pi_2$ field in an event.

\noindent
Fig. 6: Fourier power of $\pi_3$ at $\tau=4$ fm in an event.

\noindent
Fig. 7: Fourier power of $\pi_2$ at $\tau=4$ fm in an event.

\noindent
Fig. 8: $\langle A_0^1\rangle_{\rm upper}$, 
$\langle A_0^3\rangle_{\rm upper}$, and
$\langle A_0^3\rangle_{\rm lower}$ as a function
of proper time. The average is taken over 10 events.

\noindent
Fig. 9: $\langle V_0^1\rangle_{\rm upper}$, 
$\langle V_0^3\rangle_{\rm upper}$ as a function
of proper time. The average is taken over 10 events.

\noindent
Fig. 10: $\langle \pi_1\rangle_{\rm upper}$, 
$\langle \pi_3\rangle_{\rm upper}$, and
$\langle \pi_3\rangle_{\rm lower}$ as a function
of proper time. The average is taken over 10 events.

\noindent
Fig. 11: $\langle A_0^1\rangle_{\rm upper}$, 
$\langle A_0^3\rangle_{\rm upper}$, and
$\langle A_0^3\rangle_{\rm lower}$
as a function of proper time for the case with
$R_0 =5$ fm. The average is taken over 100 events.

\noindent
Fig. 12: $A_{11}(r,\tau)$ at $\tau =$ 1, 5, and 9 fm.
The average is taken over 5 events.

\noindent
Fig. 13: $A_{33}^{++}(r,\tau)$ and
$A_{33}^{+-}(r,\tau)$ at $\tau=$ 7 and 9 fm.
The average is taken over 5 events.

\noindent
Fig. 14: $A_{11}^{++}(r,\tau)$ and
$A_{11}^{+-}(r,\tau)$ at $\tau=$ 7 fm.
The average is taken over 5 events.

\noindent
Fig. 15: $B_{33}^{++}(r,\tau)$ and
$B_{33}^{+-}(r,\tau)$ at $\tau=$ 9 fm.
The average is taken over 5 events.

\noindent
Fig. 16: $B_{11}^{++}(r,\tau)$,
$C_{11}^{++}(r,\tau)$, and $C_{33}^{++}(r,\tau)$
at $\tau=$ 9 fm. The average is taken over 5 events.

\noindent
Fig. 17: Proper time evolution of the distribution of
$\pi^0$ density in an event.

\noindent
Fig. 18: Proper time evolution of the distribution of
$\pi^+$ density in an event.

\noindent
Fig. 19: Proper time evolution of the distribution
of $R(\vec{x})$.

\noindent
Fig. 20: Solid line is the proper time evolution of $R$ averaged
over 10 events. The dashed line is the same except that the momentum
integration is limited to $k< 250$ MeV.

\noindent
Fig. 21: Same as Fig. 20 except that no anomaly kick is exerted.


\begin{thebibliography}{99}

\bibitem{pw84} 
R. D. Pisarski and F. Wilczek, Phys. Rev. D {\bf 29} (1984) 338.

\bibitem{rw92}
K. Rajagopal and F. Wilczek, Nucl. Phys. {\bf B399} (1993) 395.

\bibitem{rw93} 
K. Rajagopal and F. Wilczek, Nucl. Phys. {\bf B404} (1993) 577.

\bibitem{raja97}
For a review, see: K. Rajagopal, 
in: Hwa, R.C. (ed.): {\it Quark-Gluon Plasma}, vol. 2, p. 484.

\bibitem{anselm} 
A. A. Anselm and M. G. Ryskin, Phys. Lett. {\bf B266} (1991) 482.

\bibitem{bjorken1} 
J. D. Bjorken, K. L. Kowalski, and C. C. Taylor,
SLAC Report No. SLAC-PUB-6109.

\bibitem{blaizot1}
J. -P. Blaizot and A. Krzywicki, Phys. Rev. D {\bf 46} (1992) 246.

\bibitem {centauro}
L. T. Baradzei et al., Nucl. Phys. {\bf B370} (1992) 365.
J. Lord and J. Iwai, paper submitted to International Conference
on High Energy Physics, Dallas, Texas, 1992.

\bibitem{ggp94}
S. Gavin, A. Gocksh, and R. D. Pisarski, Phys. Rev. Lett. {\bf 72}
(1994) 2143.

\bibitem{ahw95}
M. Asakawa, Z. Huang, and X. N. Wang,  Phys. Rev. Lett. {\bf 74}
(1995) 3126.

\bibitem{gl}
M. Gell-Mann and M. Levy, Nuovo Comento {\bf 16} (1960) 705.

\bibitem{rand96}
J. Randrup, Phys. Rev. Lett. {\bf 77} (1996) 1226.

\bibitem{bjorken2}
J. D. Bjorken, Phys. Rev. D {\bf 27} (1983) 140.

\bibitem{gm94}
S. Gavin and B. M\"{u}ller, Phys. Lett. {\bf B329} (1994) 486.

\bibitem{ggm93}
C. Greiner, C. Gong, and B. M\"{u}ller,
Phys. Lett. {\bf B316} (1993) 226.

\bibitem {minimax}
T. Brooks et al. (MiniMax Collaboration), hep-ph/9609375.
J. Street, Talk at Argonne Workshop on Hadron Systems at High Density
and/or High Temperature, August 7, 1997.

\bibitem {WA98}
M. M. Aggarwal et al. (WA98 Collaboration), hep-ex/9710015.

\bibitem{al96}
R. D. Amado and Y. Lu, Phys. Rev. D {\bf 54} (1996) 7075.

\bibitem{rt97}
J. Randrup and R. L. Thews, Phys. Rev. D {\bf 56} (1997) 4392.

\bibitem{kogan}
R. D. Amado and I. I. Kogan, Phys. Rev. D {\bf 51} (1995) 190.

\bibitem {mm95}
S. Mr\'owczy\'nski and B. M\"uller, Phys. Lett. {\bf B363} (1995) 1.

\bibitem{hm97}
H. Hiro-Oka and H. Minakata, TMUP-HEL-9714, hep-ph/9712476,
Phys. Lett. B, to be published.

\bibitem{ABJ} S. L. Adler,  Phys. Rev. {\bf 177} (1969) 2426;
J. S. Bell and R. Jackiw, Nuovo Cimento {\bf 60A} (1969) 47.

\bibitem{WZW}
J. Wess and B. Zumino, Phys. Lett. {\bf B37} (1971) 95;
E. Witten, Nucl. Phys. {\bf B223} (1983) 422.

\bibitem{mm96}
H. Minakata and B. M\"{u}ller, Phys. Lett. {\bf B377} (1996) 135.

\bibitem{ahw98}
M. Asakawa, Z. Huang, and X. N. Wang,  in preparation.

\bibitem{numrec}
W. H. Press, S. A. Teukolsky, W. T. Vetterling, and B. P. Flannery,
{\it Numerical Recipes} (Cambridge University Press, 1986).

\bibitem{press89} W. H. Press, B. S. Ryden, and D. N. Spergel,
Ap. J. {\bf 347} (1989) 590.

\bibitem{dah74}
G. Dahlquist and {\AA}. Bj\"{o}rck, {\it Numerical Methods}
(Prentice Hall, 1974).

\bibitem{footnote1} Hereafter we denote the conjugate fields by
$\dot{\phi}^i$ to avoid potential confusion.
\bibitem{footnote2} Since the $\pi_2$ and $\pi_3$ fields are
coupled in the equation of motion, if the time evolution of
the $\pi_3$ field is changed, that of the $\pi_2$ field is
changed as well. What we mean here is that the global behavior of
the $\pi_2$ field, such as the size of the apparent domains and
no existence of the upper-lower asymmetry, is not changed.

\bibitem{iz80}
C. Itzykson and J.-B. Zuber, {\it Quantum Field Theory} 
(McGraw-Hill, New York, 1980).

\bibitem{ammr98}
M. Asakawa, H. Minakata, B. M\"{u}ller, and K. Rajagopal,
in preparation.


\bibitem{hw94}
Z. Huang and X. N. Wang, Phys. Rev. D {\bf 49} (1994) R4335.

\bibitem{sakurai}
J. J. Sakurai, {\it Modern Quantum Mechanics} (Addison-Wesley, 1985).

\bibitem{Good75}
see e.g.: D. L. Goodstein, {\it States of Matter}, Chap. 4
(Prentice-Hall, 1975).

\bibitem{cahn64}
J. W. Cahn, J. Chem. Phys. {\bf 42} (1964) 93.

\bibitem{spinodal1}
Y. Oono and S. Puri, Phys. Rev. A {\bf 38} (1988) 434.

\end{thebibliography}
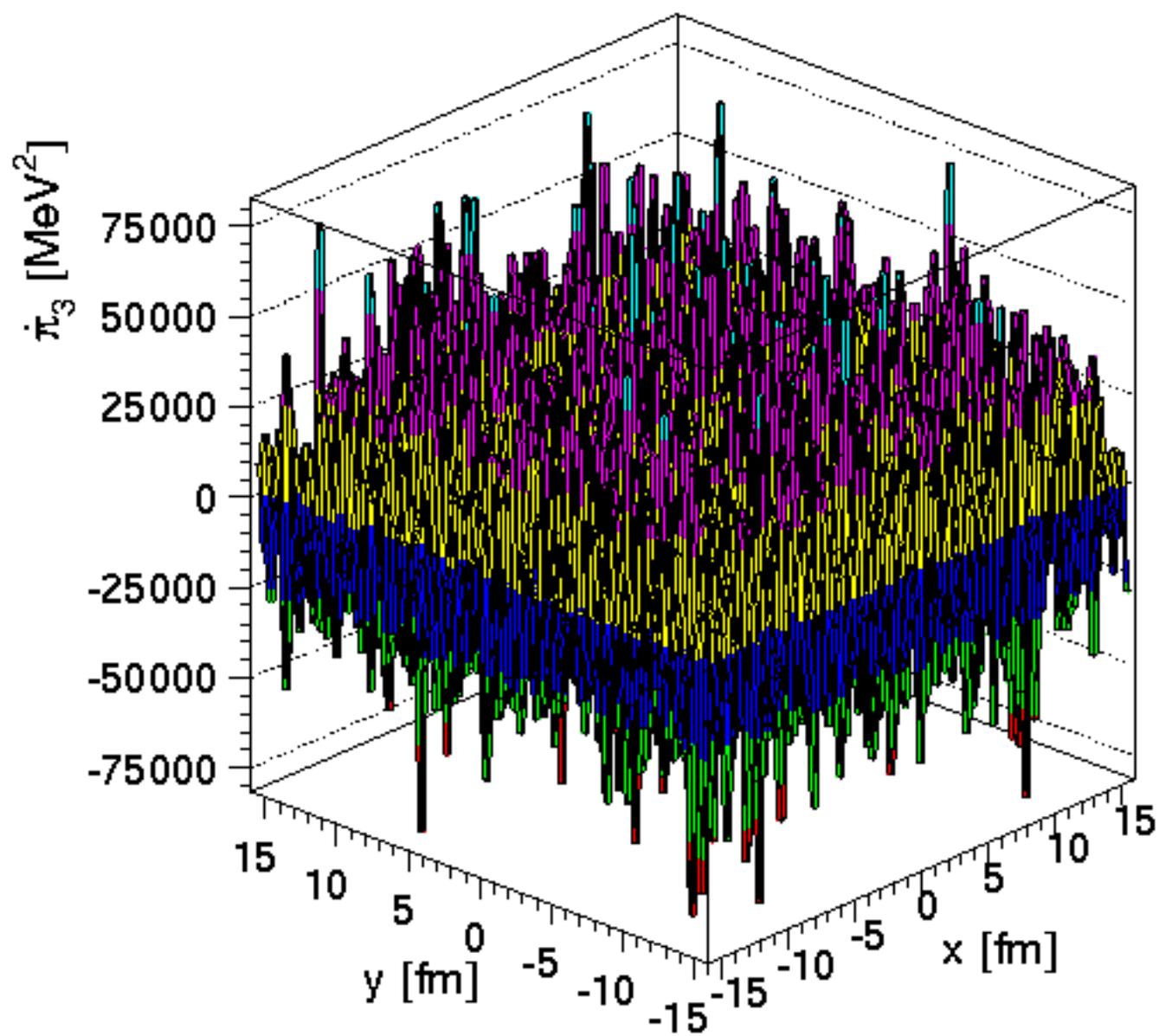
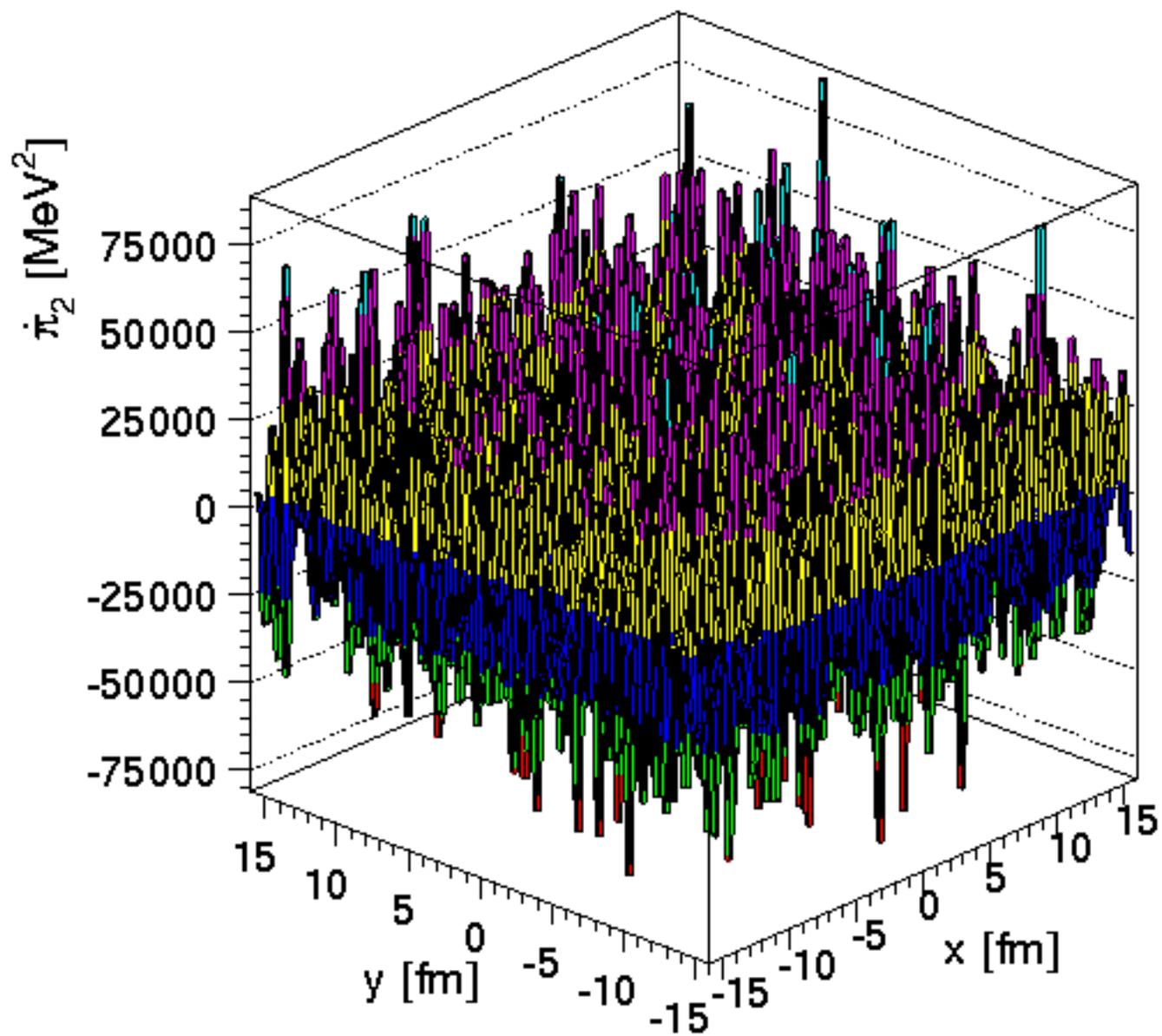
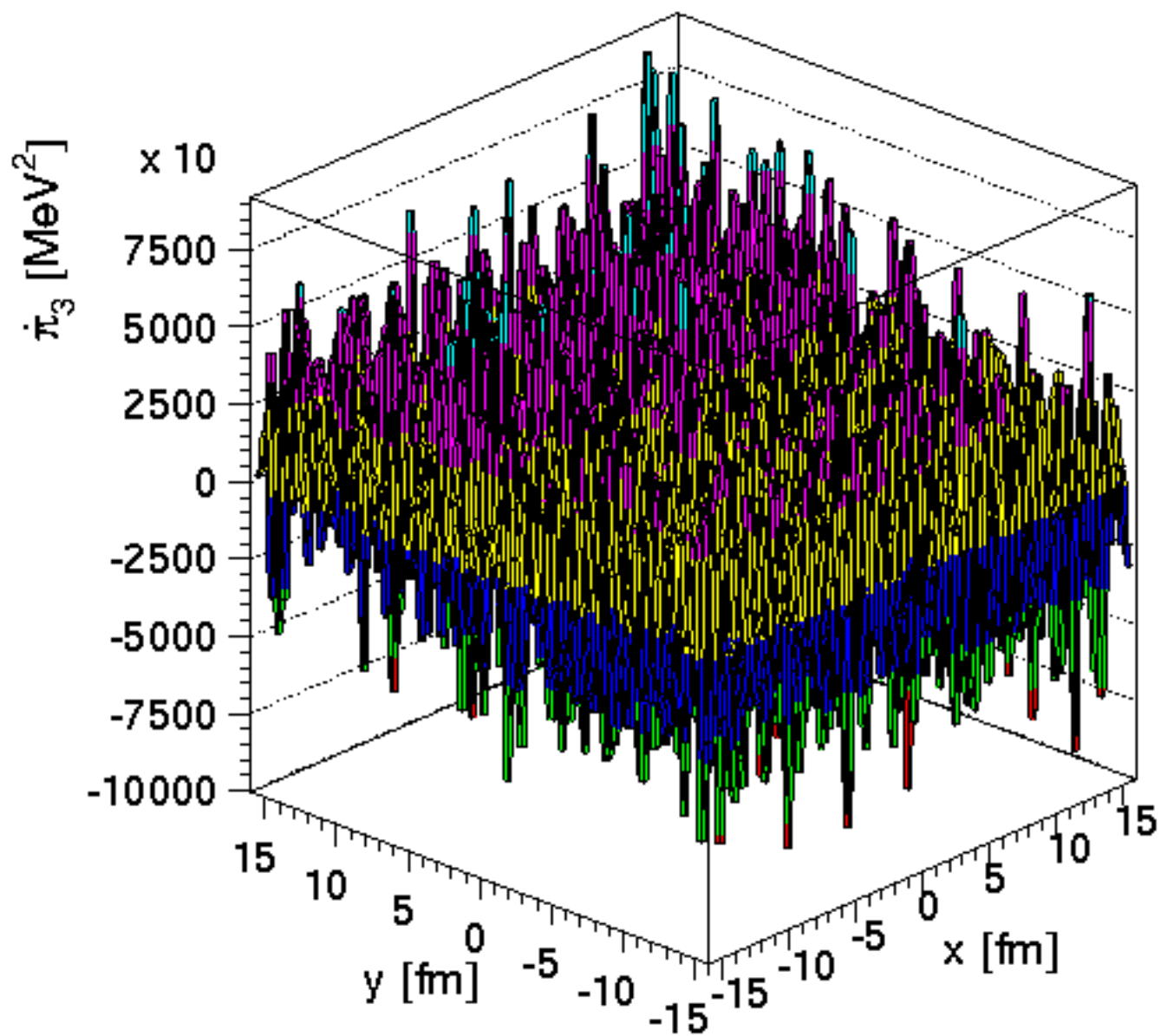
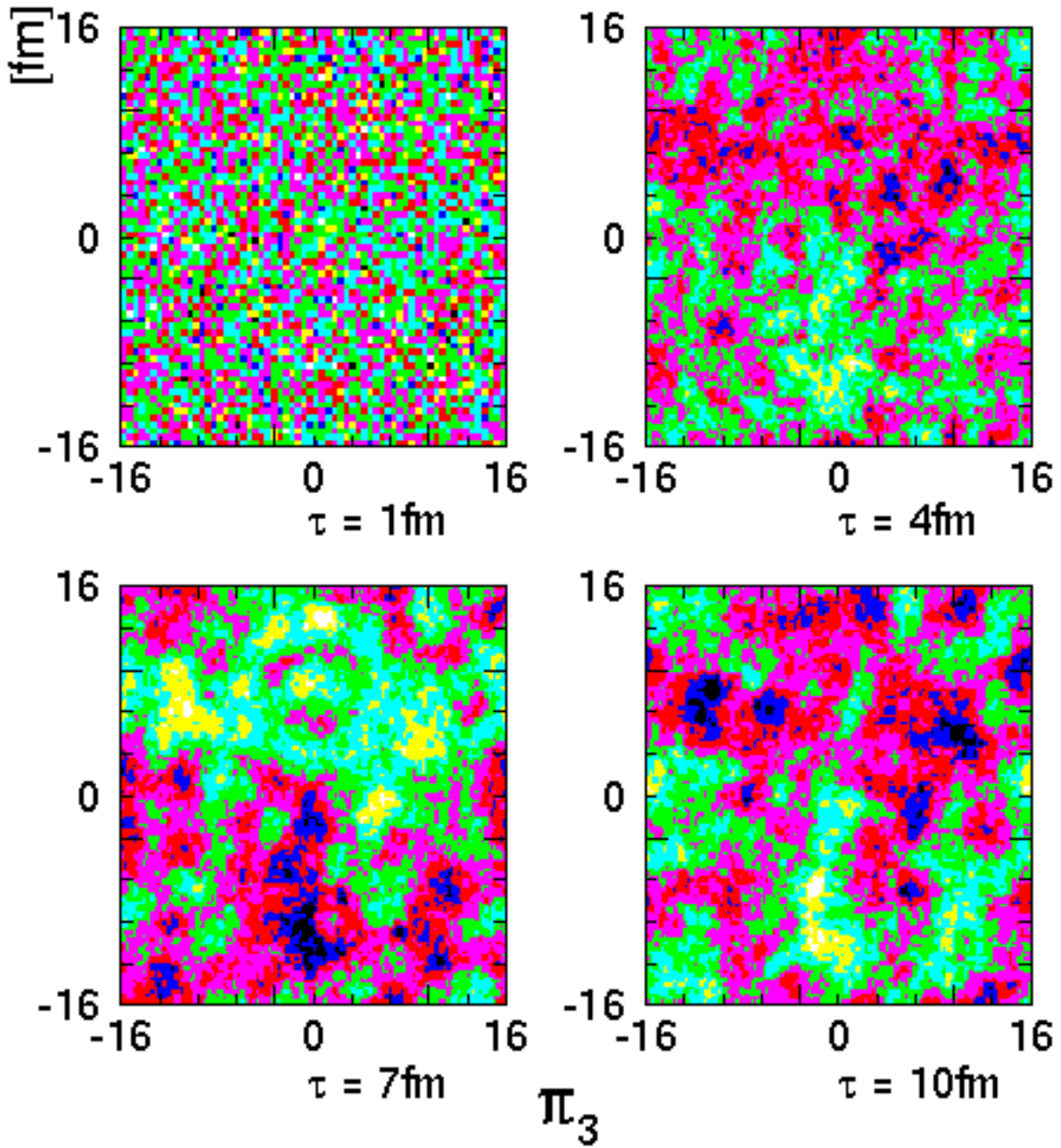
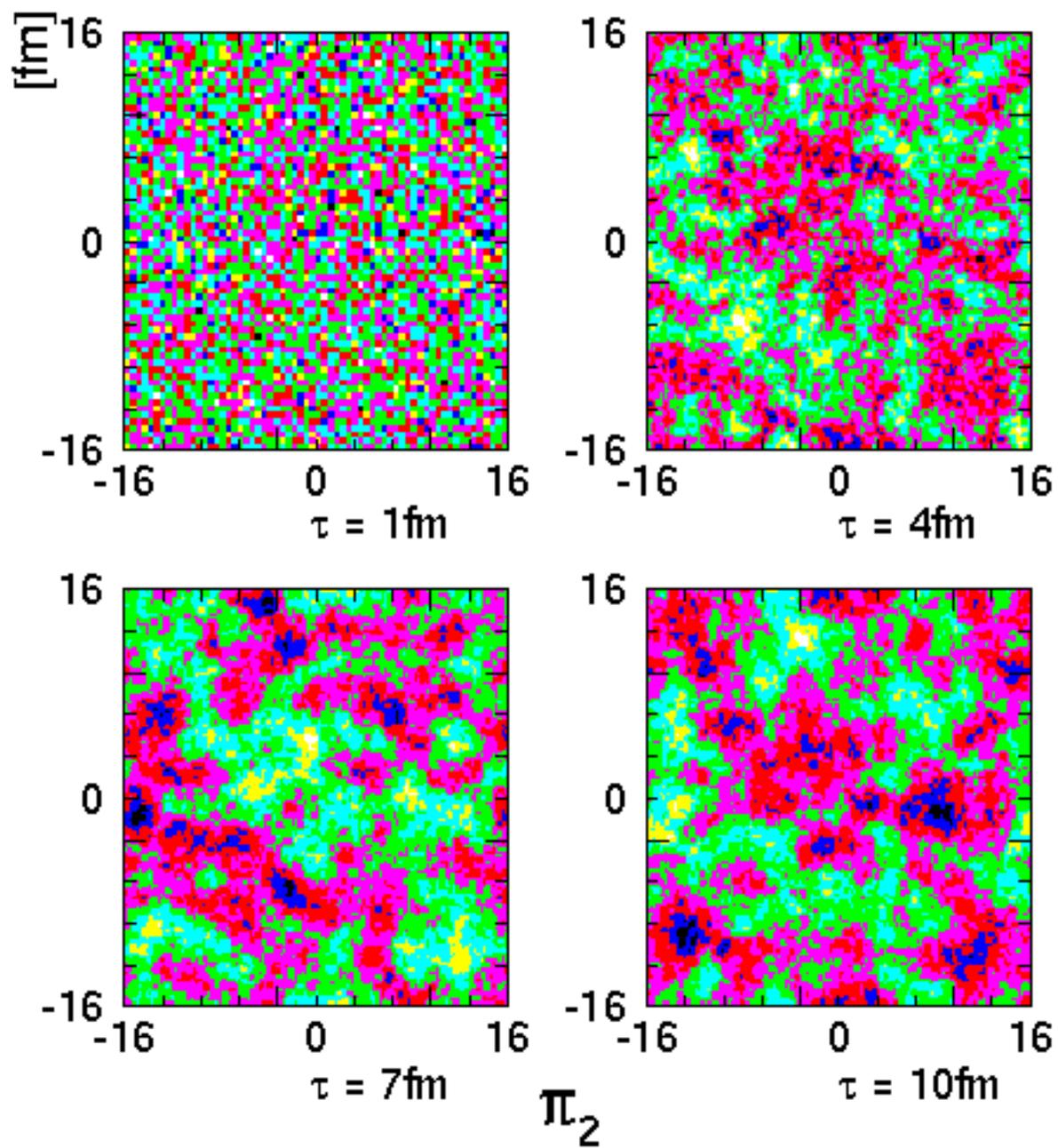
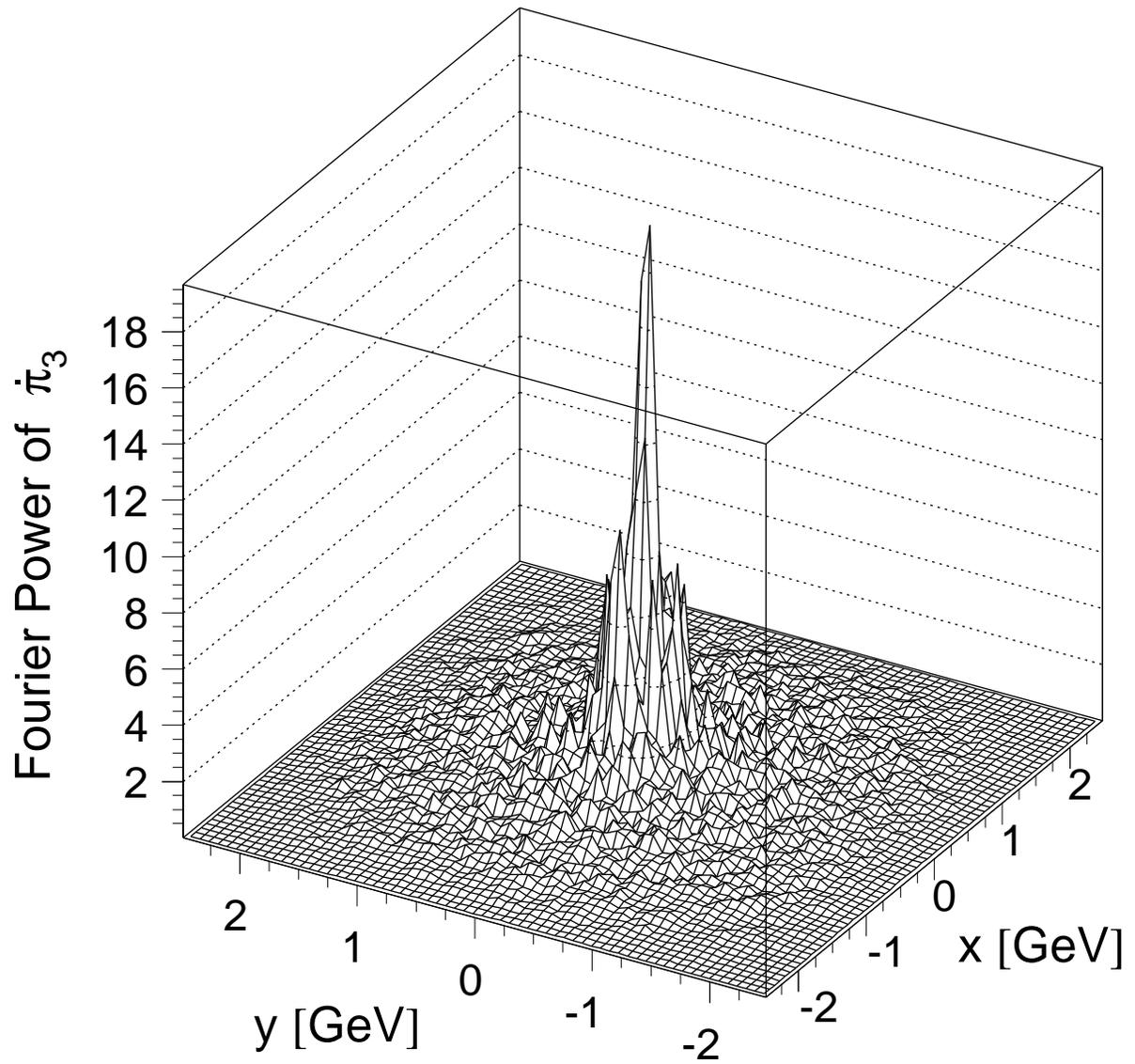
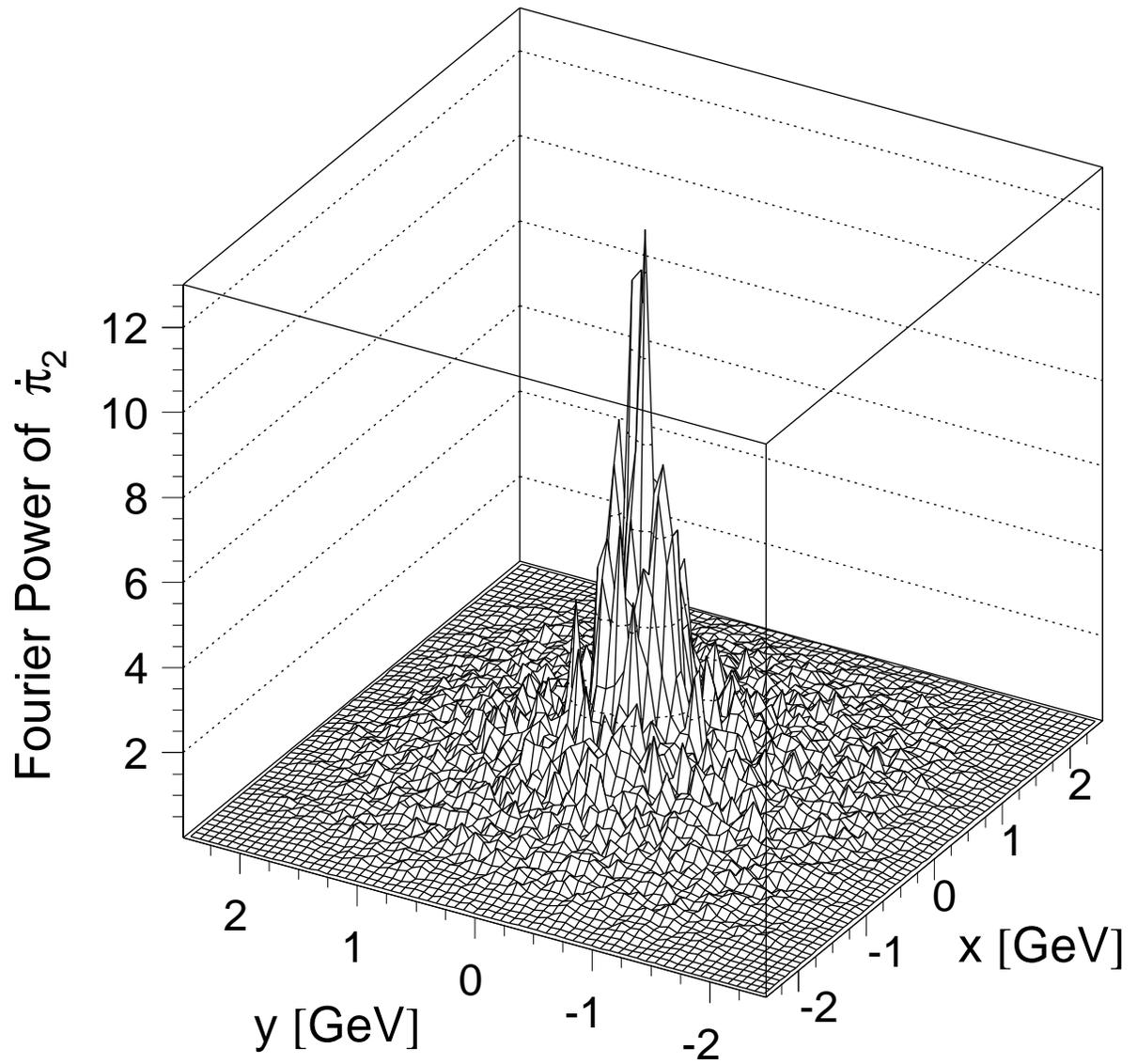
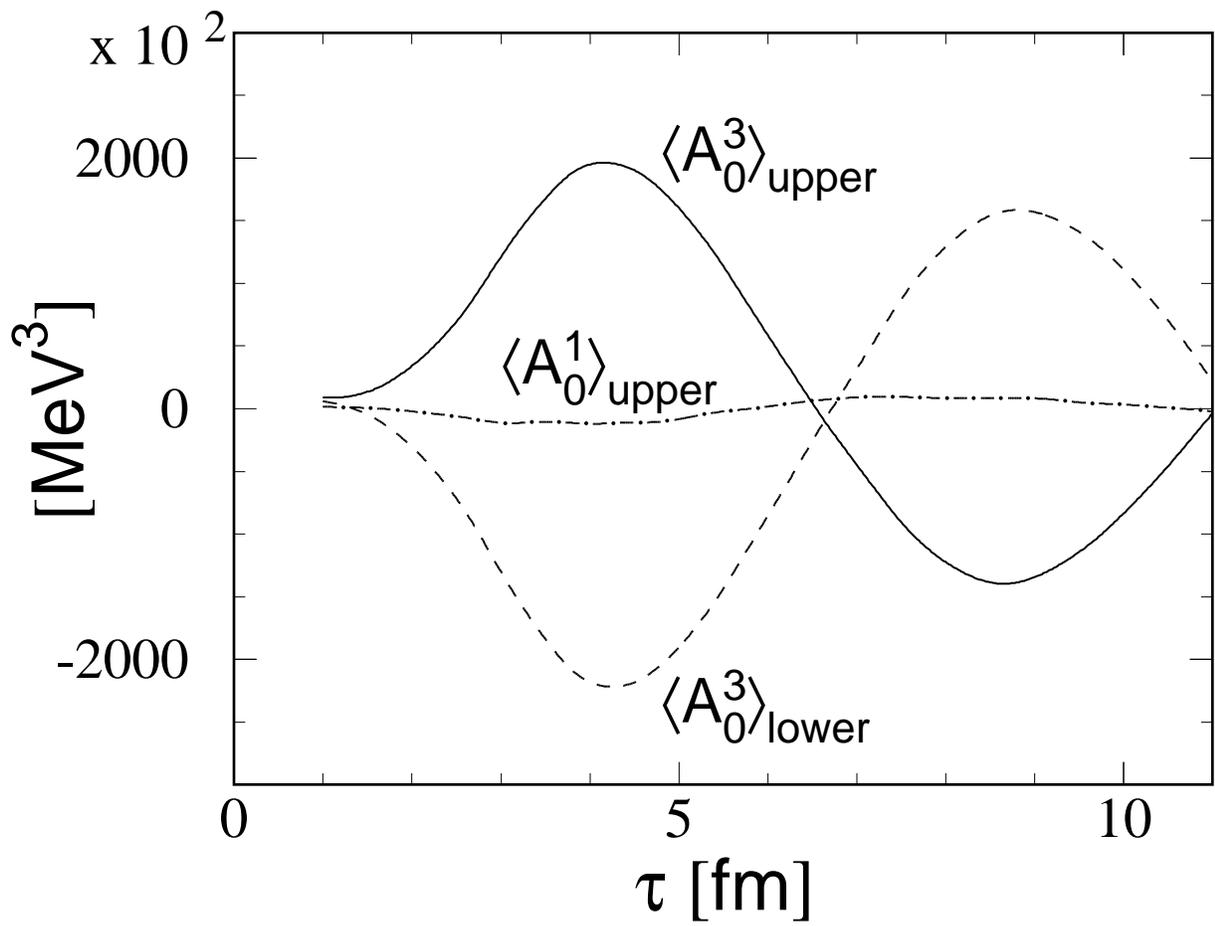
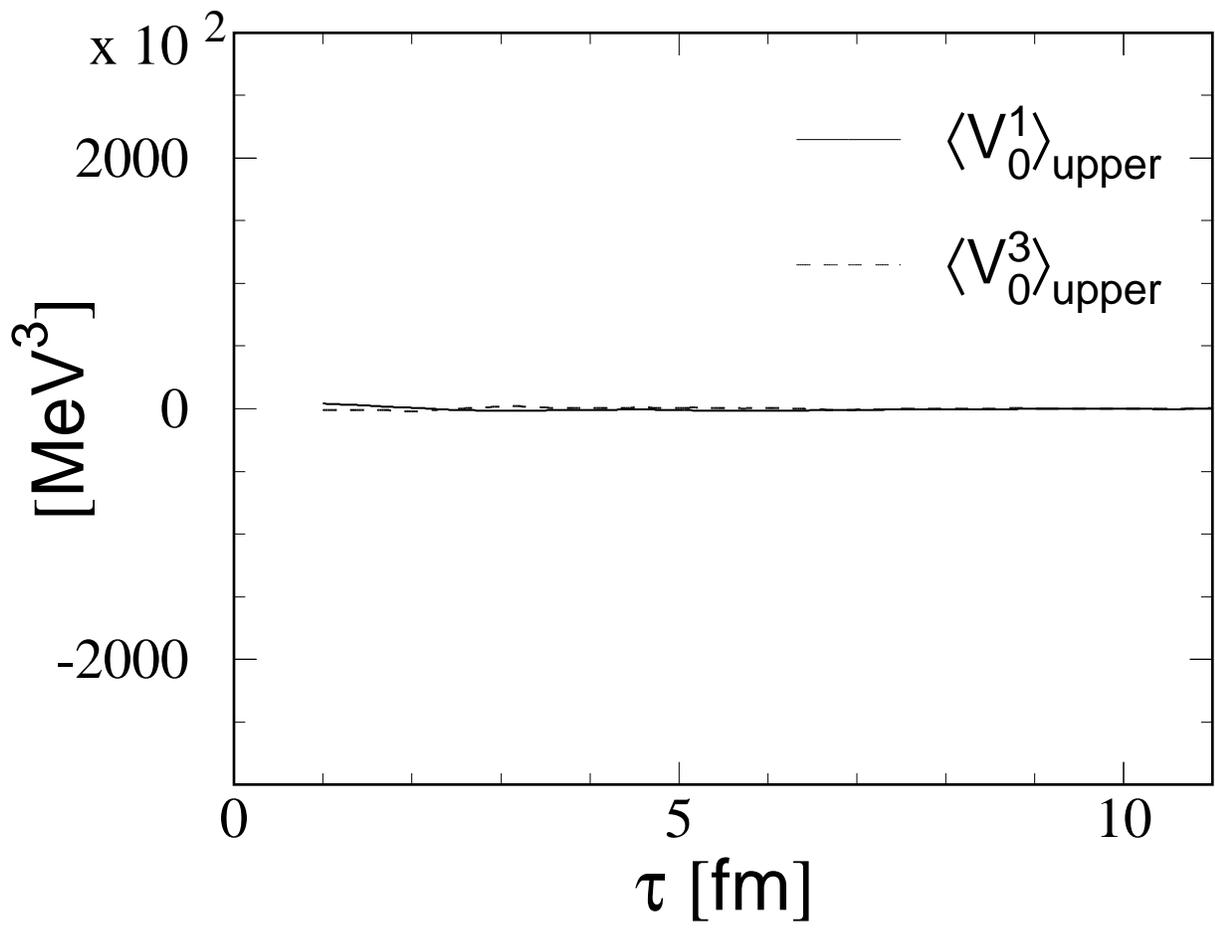
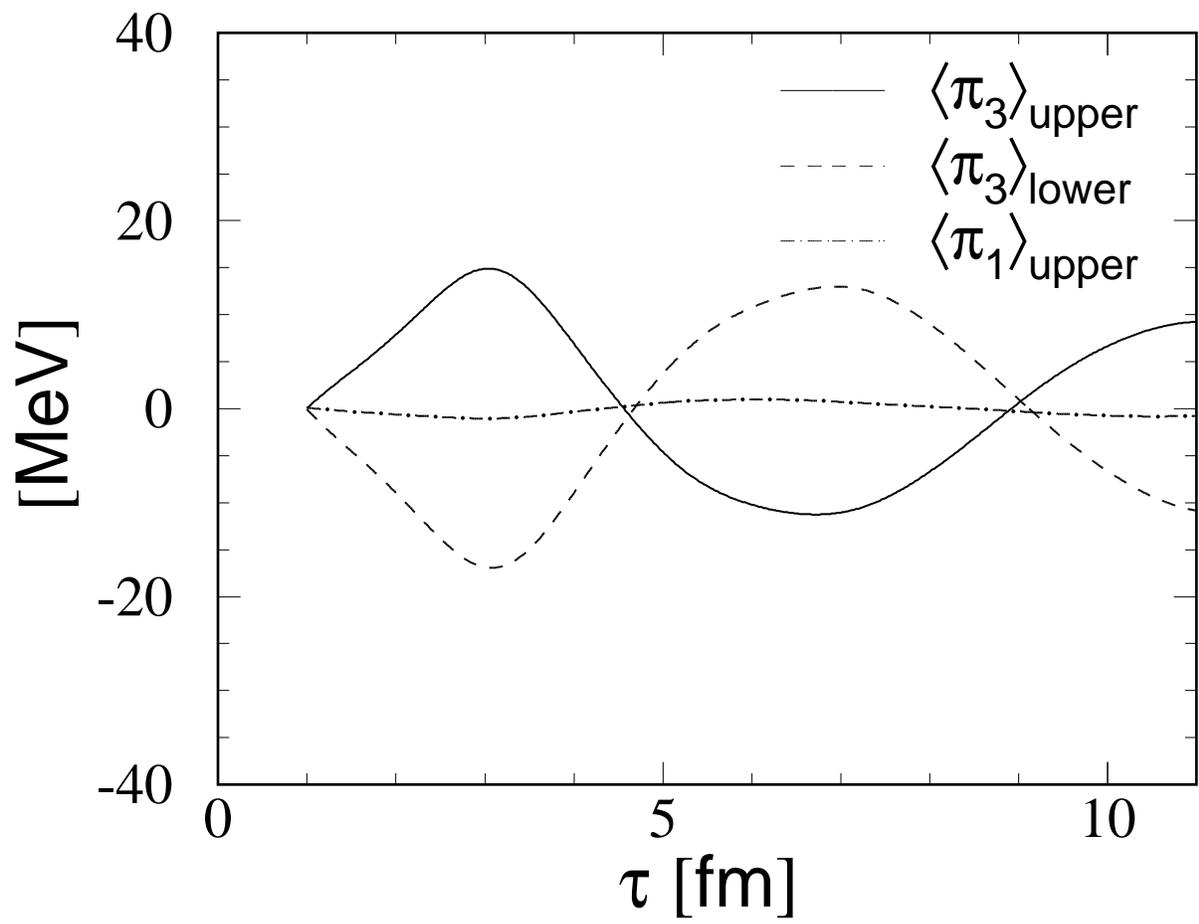
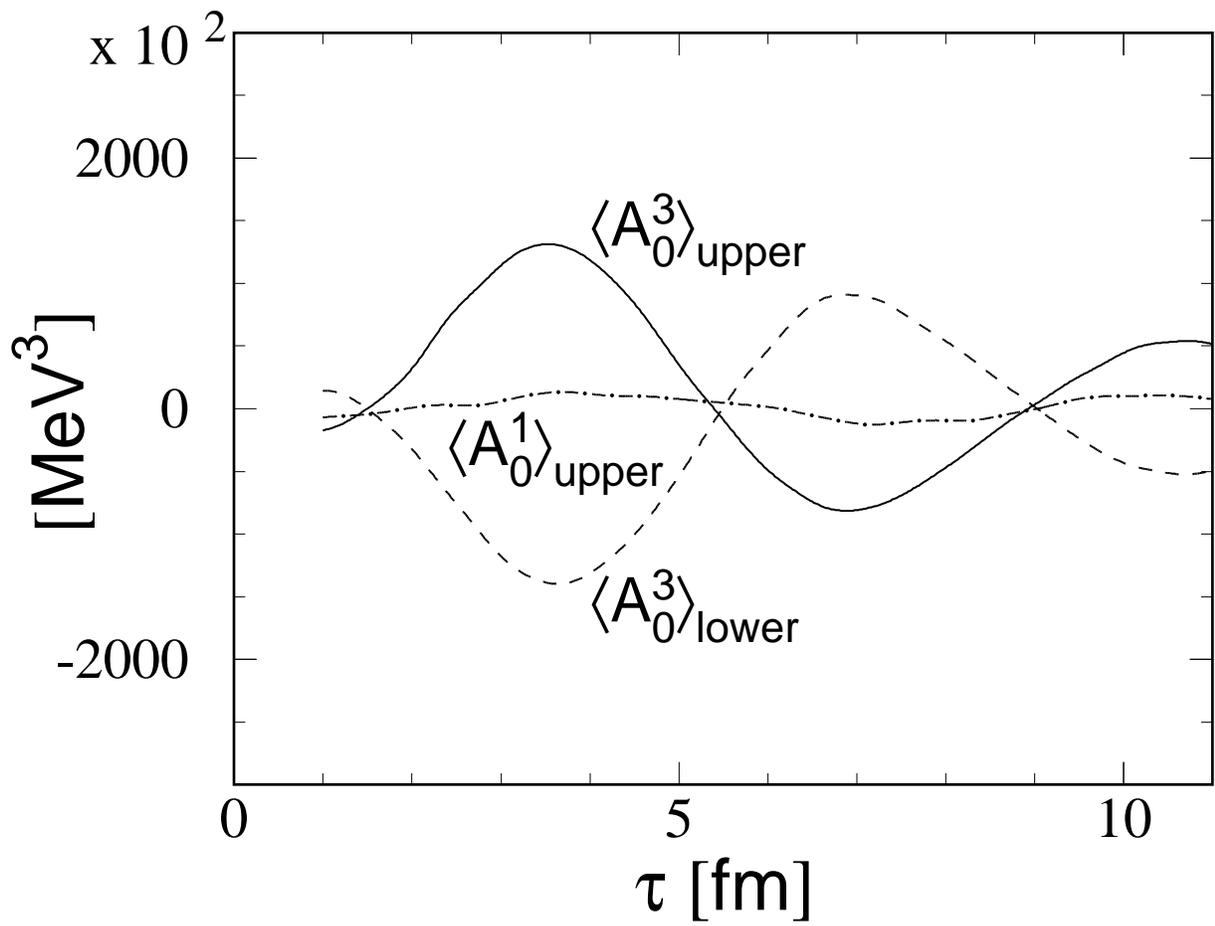
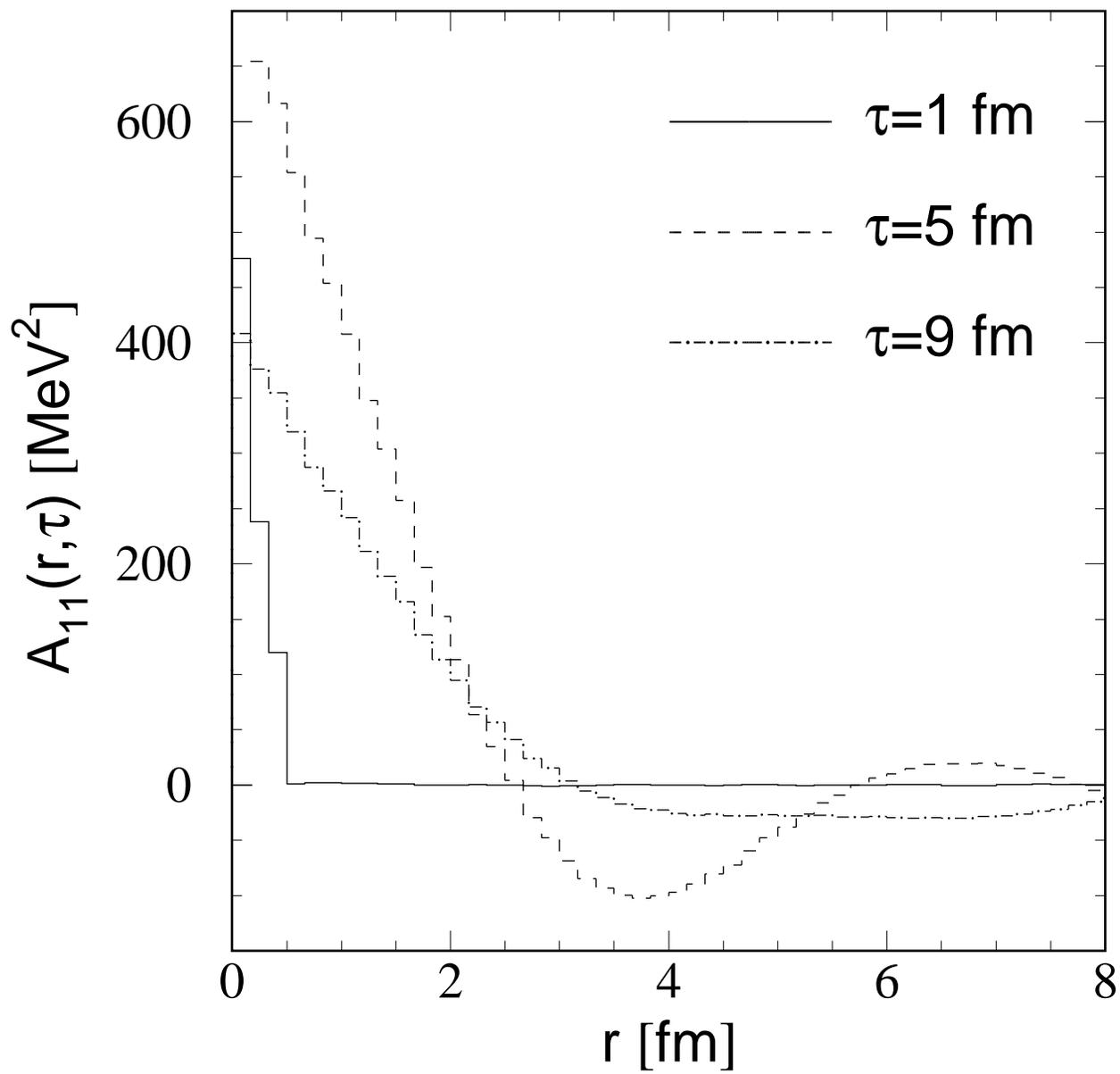
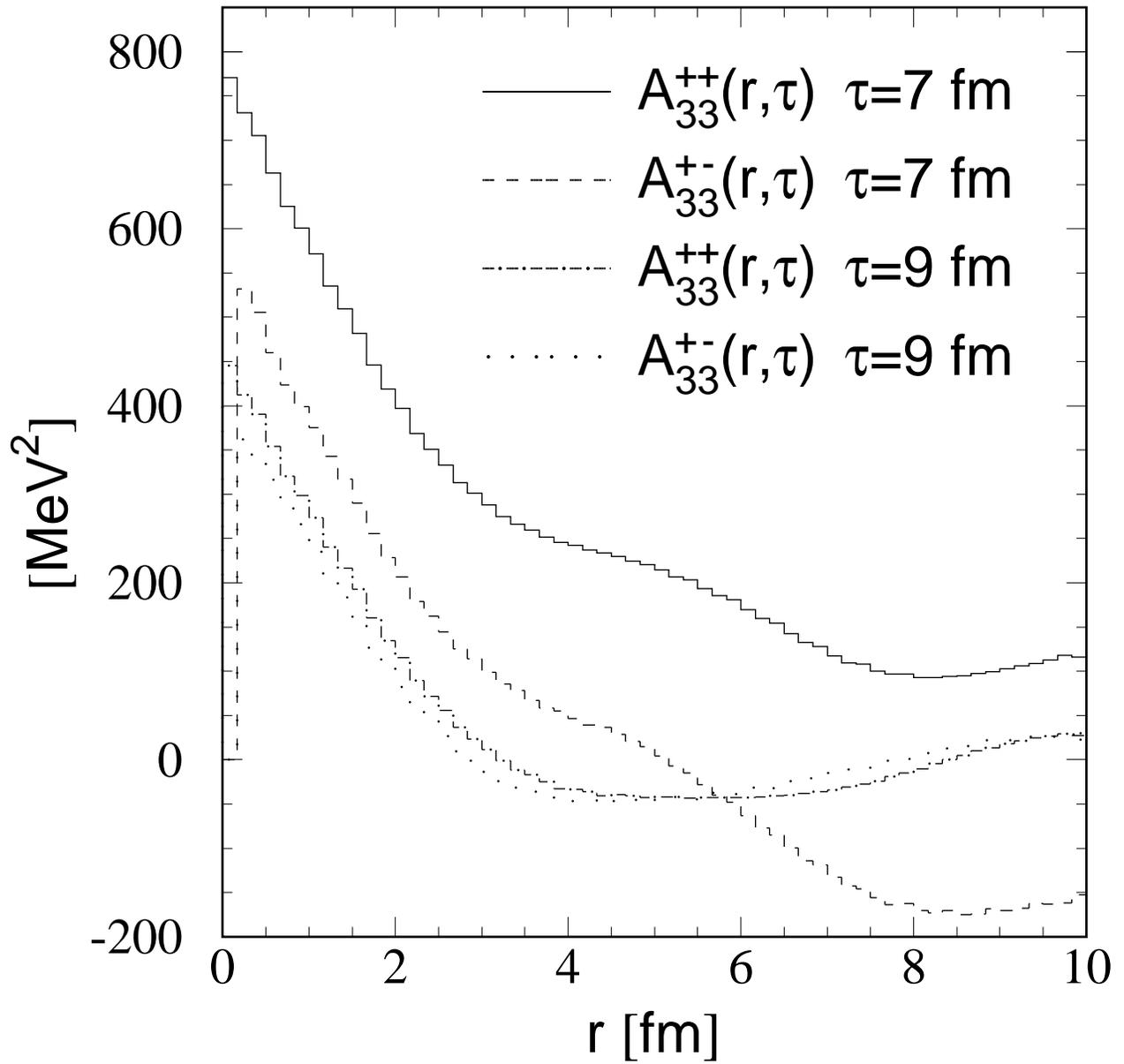
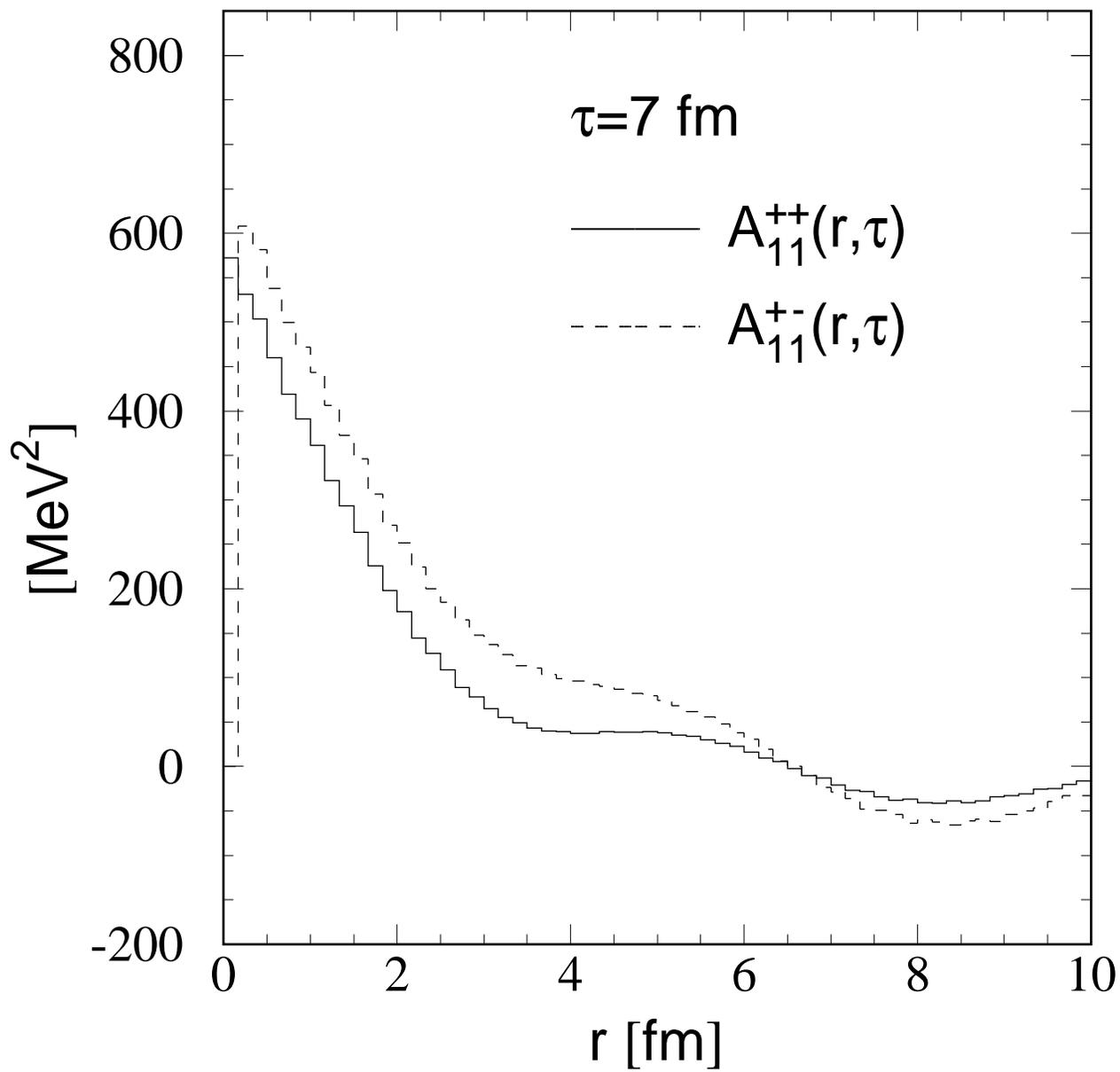
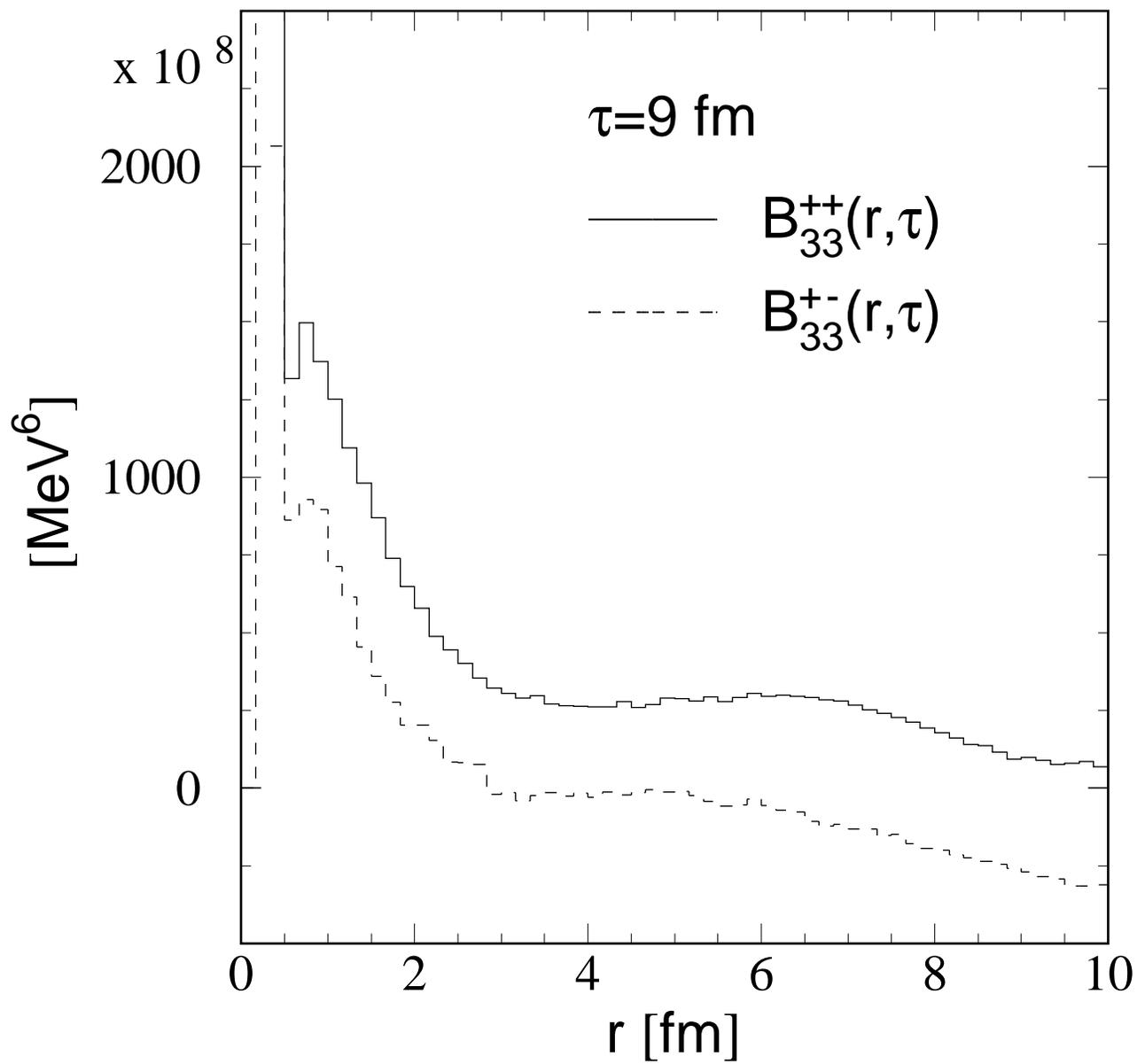
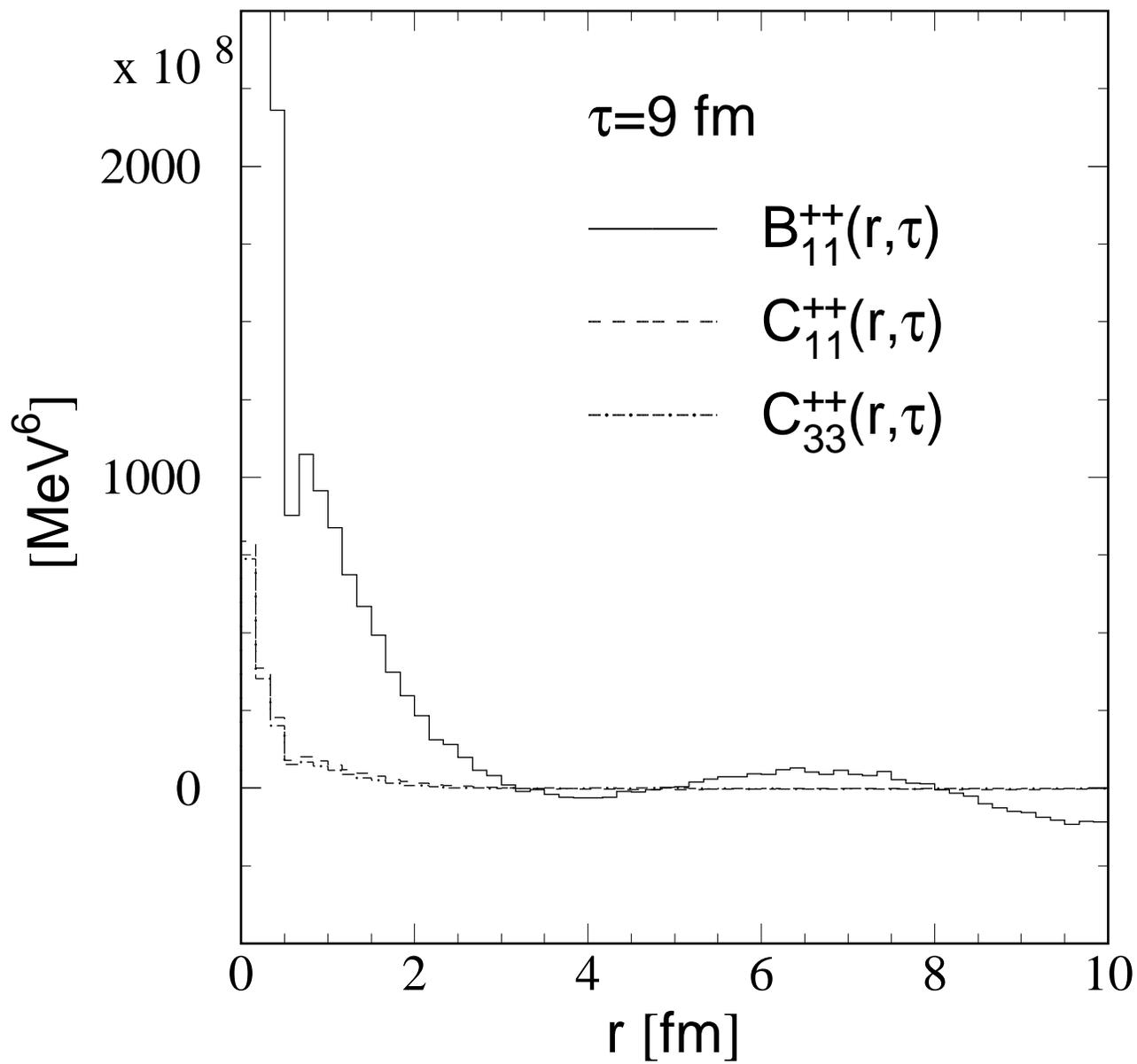
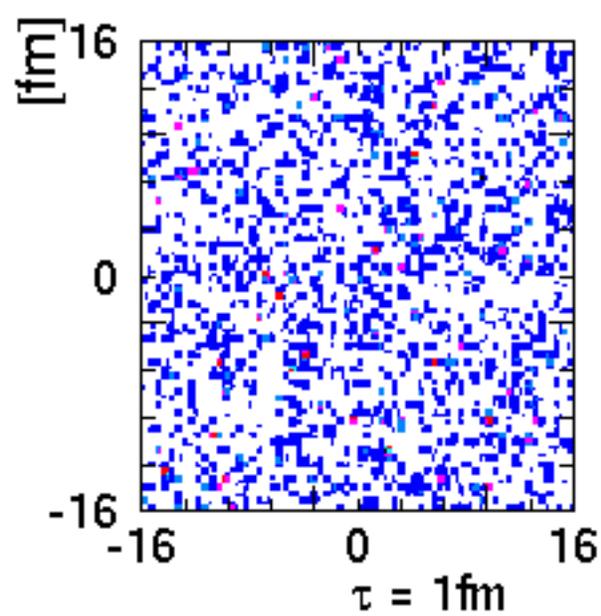
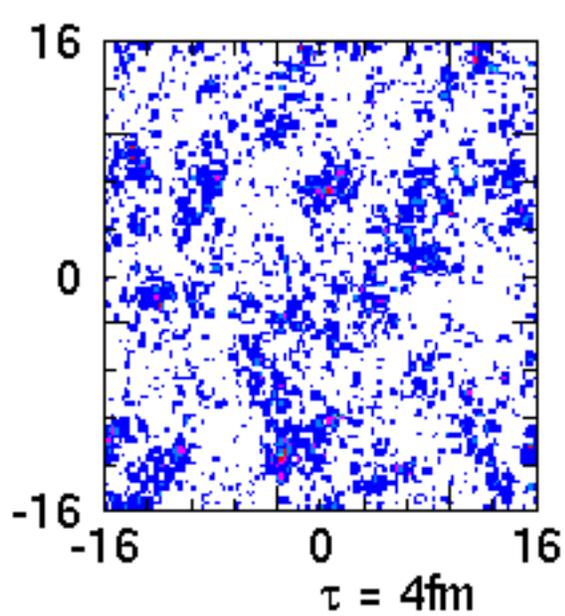
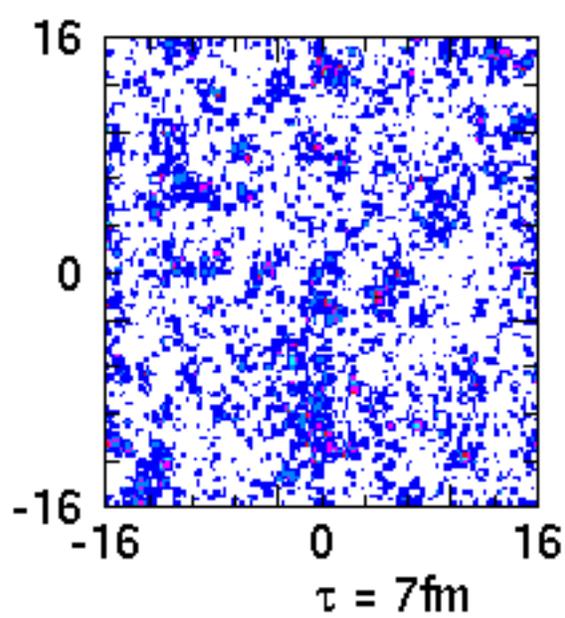
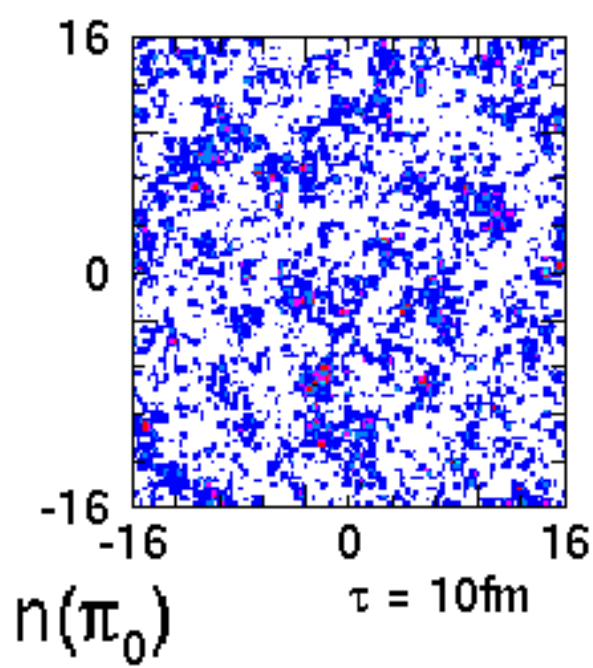
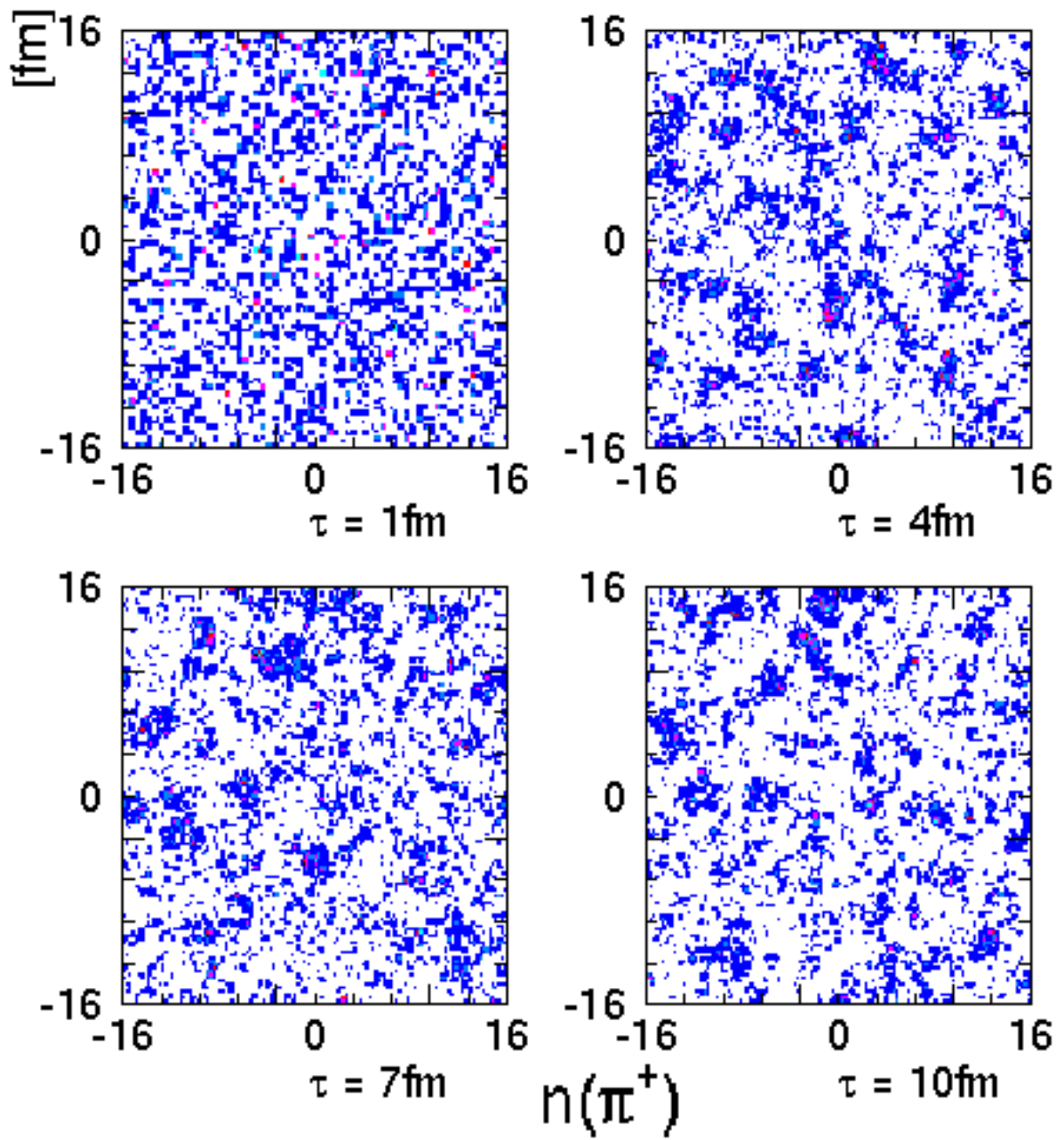
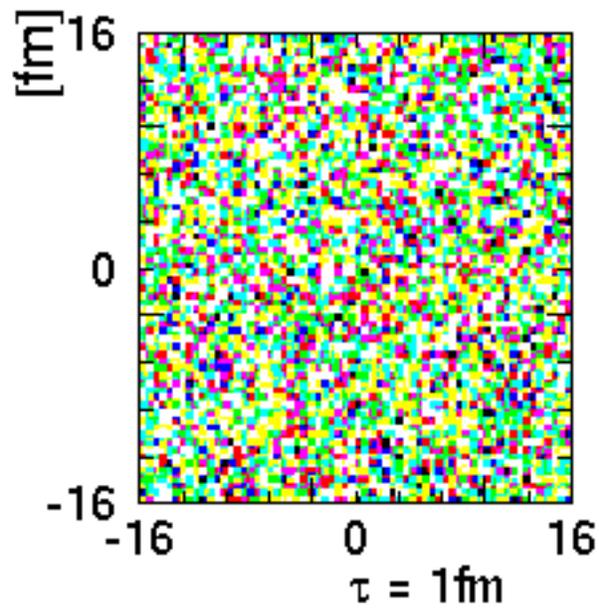
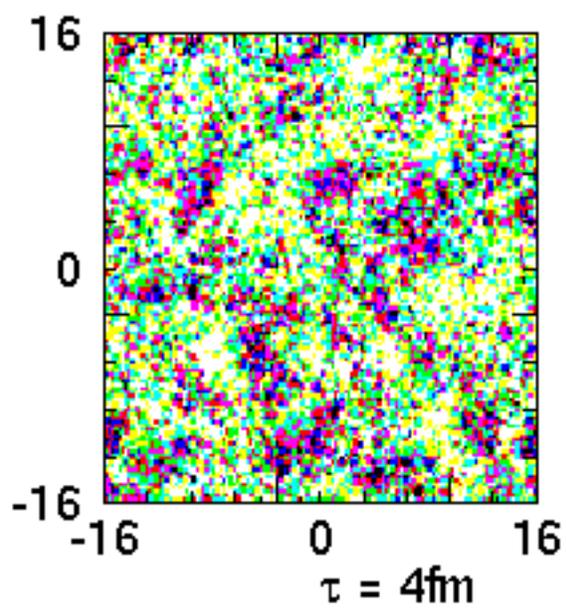
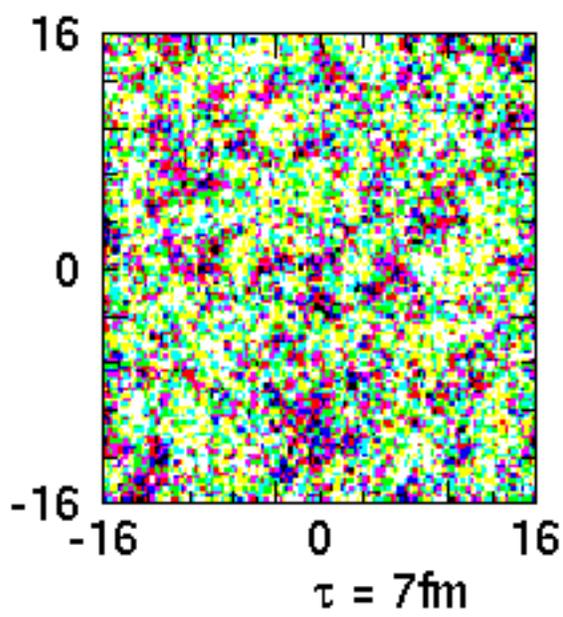
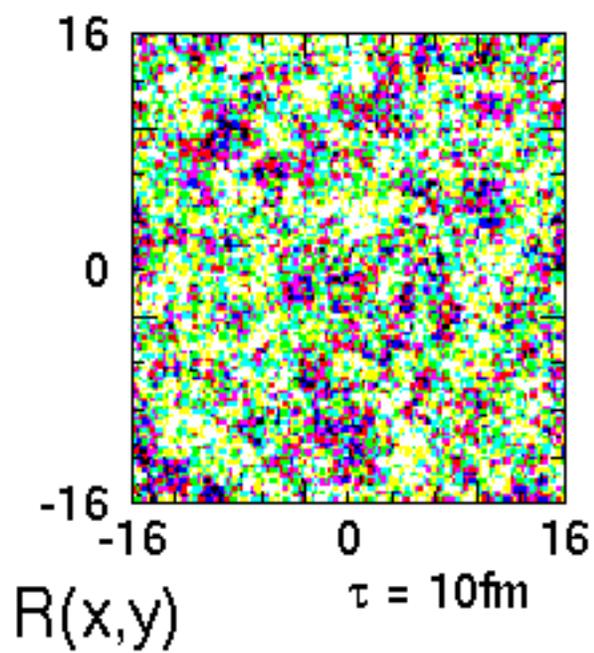
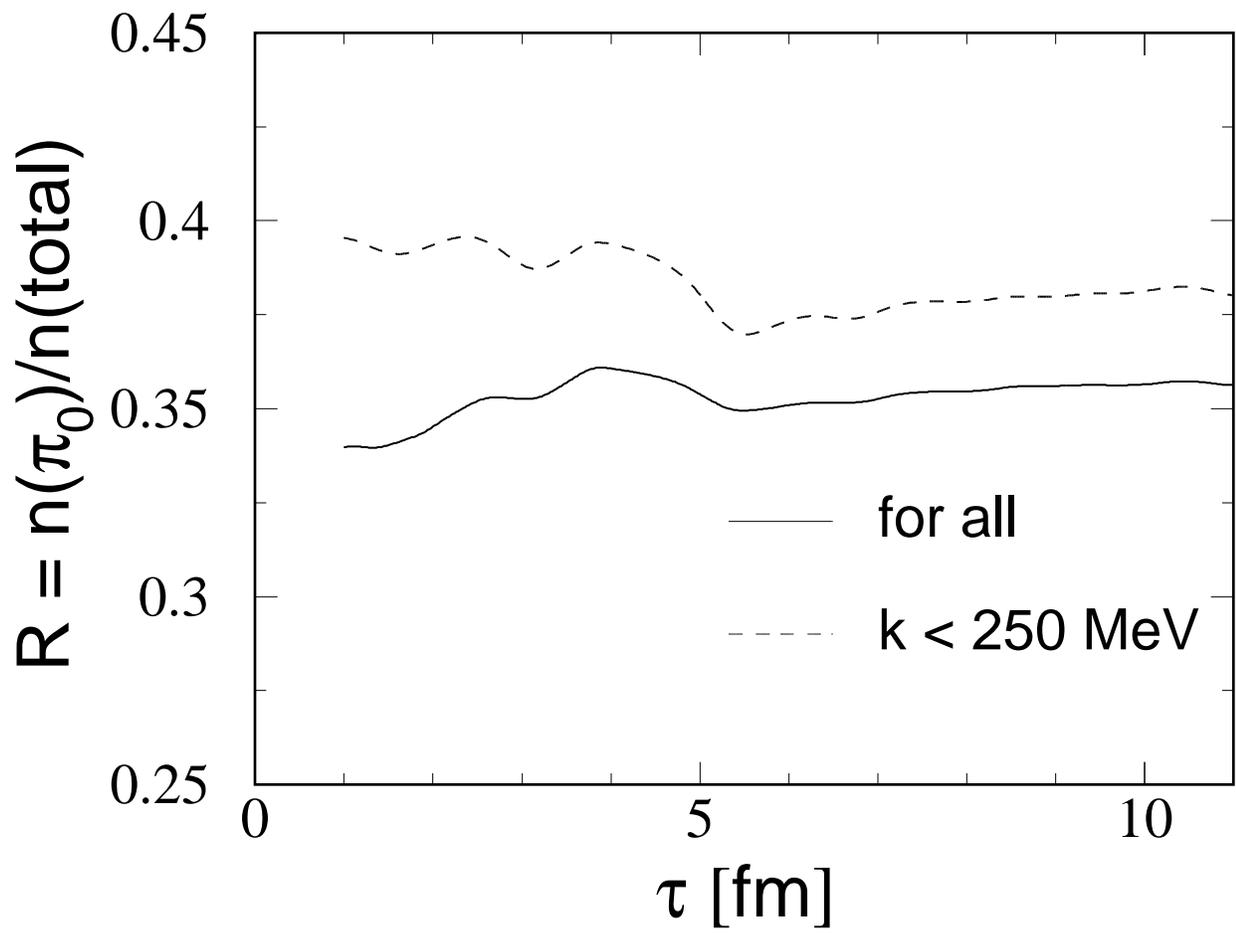
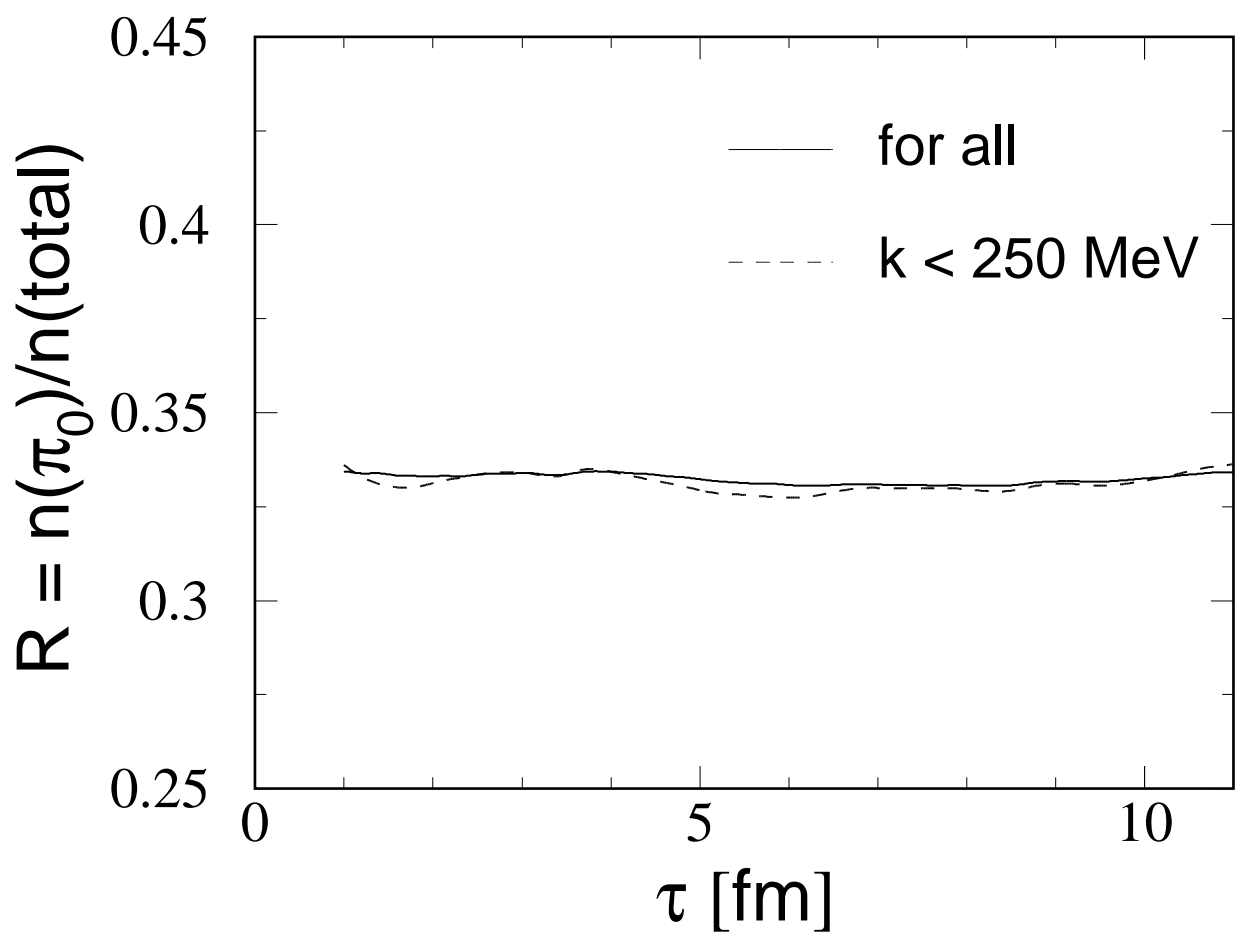
\end{document}